\documentclass[12pt,a4paper]{article}
\pdfoutput=1

\usepackage{a4wide}
\usepackage{ulem}
\usepackage{amsfonts,amssymb}
\usepackage[intlimits,fleqn]{amsmath}
\usepackage{graphicx,slashed}
\usepackage[small]{caption}
\usepackage{cite}
\usepackage[usenames,dvipsnames]{color}
     \definecolor{hgreen}{rgb}{0,.3,0}
     \definecolor{hred}{rgb}{.3,0,0}
     \definecolor{hblue}{rgb}{0,0,.3}
     \definecolor{LightGray}{gray}{0.95}
\usepackage[backref=page,
            colorlinks=true,
            linkcolor=hblue,
            citecolor=hgreen,
            filecolor=hblue,
            urlcolor=hred]{hyperref}

\numberwithin{equation}{section}

\renewcommand*{\backref}[1]{}

\providecommand{\doi}[1]{\href{http://dx.doi.org/#1}{doi:#1}}

\def\resize[#1,#2]#3{\begin{equation*}%
  \resizebox{#2\textwidth}{!}{%
  \begin{minipage}{#1\textwidth}%
  #3%
  \end{minipage}%
  }\notag%
  \end{equation*}}

\newcommand{\mk}[2]{\mkern+#1.#2mu}
\newcommand{\ol}[1]{\mkern+1mu\overline{\mkern-1mu #1 \mkern-1mu}\mkern+1mu}
\newcommand{\isum}[1]{{\textstyle\sum\limits_{#1}}}

\newcommand{\flipV}[1]{\ensuremath{\text{flip}^{V}_{f_{#1}}}}
\newcommand{\flipS}[1]{\ensuremath{\text{flip}^{S}_{f_{#1}}}}

\newcommand{\TeV}{\ifmmode\text{TeV}\else%
TeV\fi}
\newcommand{\GeV}{\ifmmode\text{GeV}\else%
GeV\fi}

\newcommand{\p}{{\phi}}
\newcommand{\ie}{{\it i.e.}\ }
\newcommand{\eg}{{\it e.g.}\ }

\newcommand{\todo}[1]{{\color{red} \ifmmode\else[todo]\fi #1}}
\newcommand{\done}[1]{{\color{green} \ifmmode\else[done]\fi #1}}
\definecolor{lightgray}{gray}{0.70}

\begin{document}
\title{
  \boldmath
    \textbf{The $Z$ Penguin in Generic Extensions of the Standard Model}
  \unboldmath
}

\author{
  $\text{Joachim Brod}^{a}$, 
  $\text{Martin Gorbahn}^{b}$\\[2em]
  {\normalsize $^{a}$Department of Physics, University of Cincinnati, Cincinnati, OH 45221, USA}\\[1em] 
  {\normalsize $^{b}$Department of Mathematical Sciences, University
    of Liverpool, Liverpool, L69 7ZL, UK}
}
\date{\today}

\maketitle

\begin{abstract}
\addcontentsline{toc}{section}{Abstract}

Precision flavour observables play an important role in the
interpretation of results at the LHC in terms of models of new
physics. We present the result for the one-loop $Z$ penguin in generic
extensions of the standard model which exhibit exact perturbative
unitarity. We use Slavnov-Taylor identities to study the implications
of unitarity on the renormalisation of the $Z$ penguin, and derive a
manifestly finite result that depends on a reduced set of physical
couplings.
\end{abstract}

\section{Introduction}

It is well known that precision flavour observables put strong
constraints on models of new physics. While these models typically
predict new particles which might be found by LHC experiments or at
future colliders, first hints could show up as anomalies in precision
flavour observables. Their examination could then lead to clues
towards the nature of new physics. It is therefore of interest to
calculate these observables in a generic extension of the standard
model (SM).

We consider an arbitrary number of additional heavy degrees of
freedom: gauge bosons, fermions and scalars. Perturbative unitarity
imposes important constraints on such generic extensions. The required
cancellation of unbounded high-energy growth of scattering amplitudes
leads to specific relations among the coupling constants that are
common to all models. These relations allow us to understand and
perform the renormalisation of the observables in a general way. The
feasibility of this approach is expected on general grounds since the
equations implied by perturbative unitarity uniquely reflect the
spontaneously broken gauge structure~\cite{LlewellynSmith:1973ey,
  Cornwall:1973tb, Cornwall:1974km} and thus may as well be derived by
means of Slavnov-Taylor identities (STI). Here we advocate the
practical implementation of those simple relations in the calculation
and renormalisation of generic loop amplitudes. This goes beyond the
typical application of perturbative unitarity in which one derives
upper bounds on yet unobserved mass spectra~\cite{Lee:1977eg,
  Chanowitz:1978uj, Chanowitz:1978mv} and combinations of masses
and/or couplings~\cite{Weldon:1984wt, Langacker:1984dma,
  Gunion:1990kf, Babu:2011sd, Grinstein:2013fia}.

As an example we study the flavour-changing transition between two SM
quarks $d_j$ and $d_i$ of different generations, induced by a heavy
neutral gauge boson at one-loop for vanishing external momenta. An
example is the FCNC $s \to d$ transition with the emission of a
virtual $Z$ boson, the so-called $Z$ penguin. It contributes, in
combination with flavor-changing box diagrams, to processes like the
rare $K \to \pi\nu\bar\nu$ decays. Just like the SM, many models of
new physics generate this transition first at the one-loop level,
since their neutral SM-fermion currents are
flavour-conserving.\footnote{Since this property drastically improves
  the potential agreement of a model with experimental flavour
  constraints, especially from $\Delta S = 2$ observables, it is
  sometimes enforced by imposing an additional $Z_2$ parity on the
  particle content (an example is $T$ parity in little Higgs
  models~\cite{Cheng:2003ju}).} In this article we discuss the
renormalisation of the general one-loop result for this process,
assuming the absence of tree-level contributions to the $d_j \to d_i$
transition. Then we provide manifestly finite results for the special
case of charged internal particles. It is then straightforward to
obtain the $Z$ penguin in any given model by just inserting the
specific couplings into our generic result. Due to sum rules derived
from the STI, only a reduced set of couplings needs to be specified in
practice. We illustrate this procedure in detail for several examples.

Our method has several interesting applications. It provides explicit
and manifestly finite results for a very general class of extensions
of the SM. The strategy is not restricted to flavour observables, but
might also be applicable, for instance, to collider and dark-matter
phenomenology.

This paper is organised as follows. After a definition of the generic
Lagrangian in Sec.~\ref{sec:generic_lagrangian}, we present in
Sec.~\ref{sec:general_result} the general analytic result which in
many models is as yet unknown, and elaborate on its
renormalisation. In Sec.~\ref{sec:specific_models} we explicitly
perform the renormalisation of our result for the case of charged
heavy particles. As an illustration, we (re-)derive the $Z$ penguin in
various renormalisable models, to wit, the SM, the two-Higgs-doublet
model, an extension of the SM with vector-like quarks, and the minimal
supersymmetric SM (MSSM). In the appendices we give the results for
the box diagrams in our notation and provide the definitions and
explicit expressions of the requisite loop functions. Moreover, we
provide the full list of Slavnov-Taylor identities for four-point
couplings.

\section{The generic Lagrangian}
\label{sec:generic_lagrangian}

In this work we consider an extension of the SM by an arbitrary number
of heavy scalar, fermion, and vector fields (in this context,
``heavy'' means that the particle masses are of the order of the
electroweak scale or larger). As our starting point we define the
parts of the generic Lagrangian which are relevant to the calculation
of box and penguin diagrams. The interaction terms involving massless
SM vector fields -- photons and gluons -- are fixed by QED and QCD
gauge invariance. In particular, the massive-massless interaction
terms are given in terms of the covariant derivative
\begin{equation} \label{eq:generic_covariant_derivative}
  (D_\mu)_{ij} = ( \partial_\mu - i e Q_F A_\mu ) \delta_{ij} - i g_s
   G_\mu^a T^a_{F,ij}
\end{equation}
by the usual kinetic terms of the massive fields $F$. Here
$T^a_{F,ij}$ and $Q_F$ generate the action of the respective gauge group
$SU(3)_c$ and $U(1)_{em}$ on the field $F$.

The three-point interactions of massive fields
\begin{equation}
  \label{eq:generic_lagrangian}
  \begin{split}
    \mathcal{L}_3 =& \mk50 \isum{f_1 f_2 s_1 \sigma} \mk50
    y_{s_1 \bar f_1 f_2}^{\sigma, abc} h_{s_1}^a
    \bar\psi_{f_1}^b P_\sigma \psi_{f_2}^c
    + \isum{f_1 f_2 v_1 \sigma} \mk50
    g_{v_1 \bar f_1  f_2}^{\sigma,abc} V_{v_1,\mu}^a
    \bar\psi_{f_1}^b \gamma^\mu P_\sigma \psi_{f_2}^c \\[2pt]
    & + \tfrac{i}{6} \mk50 \isum{v_1 v_2 v_3} \mk50
    g_{v_1v_2v_3}^{abc} \Big( V_{v_1,\mu}^a V_{v_2,\nu}^b
    \, \partial^{[\mu} V_{v_3}^{c,\nu]} + V_{v_3,\mu}^c V_{v_1,\nu}^a
    \, \partial_{\vphantom{v_2}}^{[\mu} V_{v_2}^{b,\nu]} + V_{v_2,\mu}^b
    V_{v_3,\nu}^c \, \partial_{\vphantom{v_1}}^{[\mu} V_{v_1}^{a,\nu]}
    \Big) \\[2pt]
    & + \tfrac12 \mk50 \isum{v_1 v_2 s_1} \mk50
    g_{v_1v_2s_1}^{abc} \, V_{v_1,\mu}^a V_{v_2}^{b,\mu} h_{s_1}^c
    - \tfrac{i}{2} \mk50 \isum{v_1 s_1 s_2} \mk50
    g_{v_1s_1s_2}^{abc} \, V_{v_1}^{a,\mu} \Big(
    h_{s_1}^b \, \partial_\mu h_{s_2}^c -
    \big( \partial_\mu h_{s_1}^b \big) \, h_{s_2}^c \Big)\, .
  \end{split}
\end{equation}
involve real physical scalars $h_{s_i}$, Dirac fermions $\psi_{f_i}$,
and real vector fields $V_{v_i}$, with non-zero masses $M_{s_i}$,
$m_{f_i}$ and $M_{v_i}$, respectively (complex fields are taken into
account by also including the complex conjugated field as an
independent degree of freedom in the sum, which automatically ensures
the correct normalisation). These fields are enumerated by the
corresponding indices $s_i$, $f_i$, $v_i$. The index $\sigma$ denotes
the two chiralities $\sigma = L,R$, via the chiral projectors $P_{R,L}
= (1 \pm \gamma_5)/2$. Square brackets denote antisymmetrisation of
the enclosed Lorentz indices (no symmetry factors are implied). At
one-loop level, no quartic interactions are needed for the calculation
of the $Z$ penguin. They enter, however, in the derivation of the
coupling sum rules (see Sec.~\ref{sec:STI}); the requisite Lagrangian
terms are shown in App.~\ref{sec:other_STIs}.

The sums in Eq.~\eqref{eq:generic_lagrangian} run over all particles
in a given multiplet. Consider, for instance, the last term in
Eq.~\eqref{eq:generic_lagrangian}: if $v_1$ corresponds to the SM $Z$
boson and the scalar indices to a charged scalar multiplet, the sum
runs of both positively and negatively charged scalar
particles. Alternatively, one could sum over positive particles only
and omit the factor $1/2$ in front of the sum. 

We assume that all vector fields obtain their mass by the spontaneous
breakdown of a local symmetry. The Lagrangian $\mathcal{L}_3$
comprises only the model-dependent couplings; all remaining
``unphysical'' interactions, for instance of the would-be Goldstone
bosons associated with the spontaneous symmetry breaking, can be
inferred from the requirement of perturbative unitarity, via the STIs
which we discuss below.

Due to $SU(3)\times U(1)$ gauge invariance non-vanishing couplings may
only exist for index combinations which allow the fields to combine to
form an uncharged singlet. For instance, a nonvanishing coefficient
$y_{s_1 \bar f_1 f_2}^{\sigma,abc}$ implies the charge relation
$Q_{s_1} + Q_{f_2} = Q_{f_1}$, and
\begin{align}
y^{\sigma,dbc}_{s_1 \bar
   f_1 f_2} \, T^e_{s_1, da} + y^{\sigma,abd}_{s_1 \bar f_1 f_2} \,
 T^e_{f_2, dc} = T^e_{f_1, bd} \, y^{\sigma,adc}_{s_1 \bar f_1 f_2}
 \,. \label{eq:SFF_invcond}
\end{align}
This last property is important for the calculation of QCD corrections
in the spirit of Refs.~\cite{Bobeth:1999ww, Buras:2012gm,
Buras:2012fs}. In the following, we will suppress the colour
indices. They can always be thought of as being subsumed in the field
indices $v_i$, $s_i$, and $f_i$, if necessary.

If one of the fermions (e.g., $\psi_{f_2}$) is uncharged, Schur's
lemma implies that even $T_{s_1} = T_{f_1}$.  Hermiticity puts further
restrictions on the couplings. For instance, we can express the
couplings of negatively charged Higgs and gauge bosons to fermions
through the couplings of the corresponding positively charged
particles. In general we have
\begin{equation} \label{eq:chargeflip_rules}
  \begin{gathered}
    y_{s_1 \bar f_2 f_1}^{\sigma} =
     \big( y_{\bar s_1 \bar f_1 f_2}^{\bar \sigma} \big)^* \,, \qquad
    g_{v_1 \bar f_2 f_1}^{\sigma} =
     \big( g_{\bar v_1 \bar f_1 f_2}^{\sigma} \big)^* \,, \qquad
    g_{v_1 v_2 s_1} = \big( g_{\bar v_1 \bar v_2 \bar s_1} \big)^* \,, \\
    g_{v_1 s_1 s_2} = - \big( g_{\bar v_1 \bar s_1 \bar s_2} \big)^* \,, \quad
    g_{v_1 v_2 v_3} = - \big( g_{\bar v_1 \bar v_2 \bar v_3} \big)^* \,.
  \end{gathered}
\end{equation}
The bars over bosonic indices denote the exchange of indices within a
pair of oppositely charged particles (as in $g_{\overline{W^+} \ldots}
= g_{W^- \ldots}$) and have no effect for neutral particles. The bars
over the $\sigma$s denote the opposite chirality.

We will calculate the $Z$ penguin using the general Lagrangian
Eq.~\eqref{eq:generic_lagrangian}. In practice, one would then
substitute the couplings of a given model into our final results. This
substitution is performed for several examples in
Sec.~\ref{sec:specific_models}. In particular, in this way one
recovers the SM result (see Sec.~\ref{sec:SM} for the specific
substitutions needed in this case).

\section{Slavnov-Taylor Identities for Feynman Rules}\label{sec:STI}

The constraints derived from perturbative unitarity reflect a
spontaneously broken gauge symmetry. To exploit these constraints for
our generic Lagrangian we use the STIs of an arbitrary fundamental
spontaneously broken gauge theory. The massive vector fields of
\eqref{eq:generic_lagrangian} are the gauge bosons of the fundamental
theory supplemented by a standard $R_\xi$ gauge-fixing term. This has
two consequences. First, the couplings of Goldstone bosons can be
linked directly to the couplings of the corresponding vectors in the
mass-eigenstate basis. This use of STIs is well known and summarized
in the Goldstone-boson equivalence theorem \cite{Cornwall:1974km,
  Lee:1977eg, Chanowitz:1985hj, Gounaris:1986cr}. Second, we obtain
certain sum rules, \ie equations that impose non-trivial constraints
on the couplings of physical fields and encode the full spontaneously
broken gauge structure on the level of Feynman rules\footnote{The
  couplings in \eqref{eq:generic_lagrangian} are defined such that the
  Feynman rules are given, after multiplication by a factor of $i$, in
  terms of the usual Lorentz structures in the conventions of {\tt
    FeynArts} \cite{Hahn:2000kx}.}. We will use these sum rules later
to renormalise the $Z$ penguin.

From a technical point of view it is easiest to derive the sum rules
from the vanishing Becchi-Rouet-Stora-Tyutin (BRST)
\cite{Becchi:1975nq,Tyutin:1975qk} transformation of suitable vertex
functions. To start, we note that throughout this work the gauge
freedom of~\eqref{eq:generic_lagrangian} is fixed with a standard
linear $R_\xi$ Lagrangian~\cite{Fujikawa:1972fe}
\begin{equation}\label{eq:gaugefix_massive_vector}
\mathcal{L}_\text{fix} = - \sum_v (2 \xi_{v})^{-1} F_{\bar v} F_v\,,
\qquad F_v = \partial_\mu V_v^\mu - \sigma_v \xi_v M_v \phi_v \,,
\end{equation}
for every vector field $V_v$ of mass $M_v$ and the corresponding
pseudo-Goldstone boson field $\phi_v$. The coefficients $\sigma_v$ can
have the values $\pm i$ for complex fields and $\pm 1$ for real
fields. For the SM fields they are given by $\sigma_{W^\pm} = \pm i$
and $\sigma_Z = 1$, and we choose this convention in general for all
charged and neutral vector fields.

By applying a BRST transformation $s$ to a Green's function
\begin{equation}
G^{\bar{u}_{v} (\ldots)_\text{ph}} \equiv \left \langle T \big\{
\bar{u}_{v} (\ldots)_\text{ph} \big\} \right \rangle
\end{equation}
which involves an anti-ghost field $\bar{u}_{v}$, and using the
transformation property $s \bar{u}_{v} = - F_v/\xi_v$, we obtain a
linear relation between the connected, truncated Green's functions
$G_c^{\underline{W_v} (\ldots)_\text{ph}}$ and
$G_c^{\underline{\phi_v} (\ldots)_\text{ph}}$. Here, the dots
$(\ldots)_\text{ph}$ stand for any combination of physical, asymptotic
on-shell fields, whose BRST variations vanish. The underlining of a
field indicates that the corresponding external leg has been
amputated. In our convention, labels on vertex functions denote
outgoing fields, whereas all momenta are incoming\footnote{The
  momentum configuration of the vectors and Goldstone bosons appearing
  in the gauge-fixing function is not restricted any further, in
  contrast to the procedure used, for instance, in applications of the
  Goldstone-boson equivalence theorem~\cite{He:1994br}.}. Angle
brackets denote a vacuum expectation value, and $T \{ \ldots \}$ the
time-ordered product of fields.

The STIs lead to the following relation in momentum space:
\begin{equation}
  \begin{pmatrix} k^\mu \\ i \sigma_{\bar v} \xi_v M_v
  \end{pmatrix}^T G^{\bar v}_{(\mu\nu)} \begin{pmatrix} \langle
    \underline{V}_v^{\nu} (\ldots)_\text{ph} \rangle_c \\ \langle
    \underline{\phi}_v (\ldots)_\text{ph} \rangle_c
  \end{pmatrix} = 0 \,.
\end{equation}
$G^{\bar v}_{(\mu\nu)}$ denotes the propagator
for a vector boson or its Goldstone boson, and is given by the inverse
of the two-point vertex function $\Gamma^{v\,(\mu \nu)}$. These
functions can be decomposed as
\begin{equation} \label{eq:gf_and_vf_are_inverse}
     \Gamma^v_{(\mu\nu)}(k, -k) = \begin{pmatrix}
      \isum{P=T,L} g^P_{\mu\nu} \Gamma^{V_v V_{\bar{v}}}_P(k^2)
      & k_\mu \Gamma^{V_v \phi_{\bar{v}}}_L(k^2) \\
      k_\nu \Gamma^{\phi^a V_{\bar{v}}}_L(k^2)
      & \Gamma^{\phi_v \phi_{\bar{v}}}(k^2) \end{pmatrix} \,, \quad
    G^{\bar{v}}_{(\mu\lambda)} \Gamma^{v(\lambda \nu)} = i
    \begin{pmatrix} \delta_\mu^\nu & 0 \\ 0 & 1 \end{pmatrix} \,.
\end{equation}
where $g^T_{\mu\nu} \equiv g_{\mu\nu} - \frac{k_\mu k_\nu}{k^2}$ and
$g^L_{\mu\nu} \equiv g_{\mu\nu} - g^T_{\mu\nu}$. It
follows~\cite{Yao:1988aj} that the STIs are given by
\begin{equation}
  \label{eq:st_identity}
  \langle T \Big\{ k^\mu \underline{V_v^{\mu}} - i
  \sigma_{\bar v} M_v A_v(k^2) \underline{\phi_v} \Big \}
  (\ldots)_\text{ph} \rangle = 0 \,, \quad
  A_v = \frac{\Gamma^{V_v V_{\bar v}}_L + \frac{k^2}{\xi_v}}{M_v
    \big(M_v - i \sigma_v \Gamma^{V_v \phi_{\bar v}}_L\big)} \,.
\end{equation}
In principle, one would have to account for the mixing of different
bosons; consider, for instance, $Z - A$ mixing in the SM.  However,
this only affects Eq.~\eqref{eq:st_identity} at loop level, while we
use the equation to evaluate Feynman rules and tree-level sum rules,
where, in fact, $A_v(k^2) = 1$. Evaluating this identity at tree level
shows that the three-point couplings involving Goldstone bosons
$\phi_v$ are related to the couplings involving the corresponding
gauge bosons $V_v$ as follows:
\begin{equation} \label{eq:STI_3pt_GB_replacements}
  \begin{gathered}
    \begin{aligned}
      & g_{v_1\phi_2\phi_3} = \sigma_{v_2} \sigma_{v_3} \,
      \tfrac{M_{v_2}^2 + M_{v_3}^2 - M_{v_1}^2}{2 \,
        M_{v_2} M_{v_3}} \, g_{v_1v_2v_3} \,,&\;
      & g_{\phi_1\phi_2s_1} = - \sigma_{v_1} \sigma_{v_2} \,
      \tfrac{M_{s_1}^2}{2 \, M_{v_1} M_{v_2}} \,
      g_{v_1v_2s_1} \,,\\
      & g_{v_1v_2\phi_3} = - i \sigma_{v_3} \,
      \tfrac{M_{v_1}^2 - M_{v_2}^2}{M_{v_3}} \,
      g_{v_1v_2v_3} \,,&\;
      & g_{\phi_1s_1s_2} = i \sigma_{v_1} \,
      \tfrac{M_{s_1}^2 - M_{s_2}^2}{M_{v_1}} \,
      g_{v_1s_1s_2} \,,\\
      & g_{v_1\phi_2s_1} = - i \sigma_{v_2} \, \tfrac{1}{2
        \, M_{v_2}} \, g_{v_1v_2s_1} \,,&\;
      & g_{\phi_1\phi_2\phi_3} = 0 \,,\\
    \end{aligned} \\
    y_{\phi_1 \bar f_1 f_2}^{\sigma}= - i \sigma_{v_1} \,
    \tfrac{1}{M_{v_1}} \, \big( m_{f_1} g_{v_1 \bar f_1 f_2}^{\sigma}
    - g_{v_1 \bar f_1 f_2}^{\bar \sigma} m_{f_2} \big) \,.
  \end{gathered}
\end{equation}
Here the subscripts $\phi_i$ correspond to Goldstone-boson indices and
are used in distinction to the subscripts $s_i$ corresponding to
physical scalars (for instance, Higgs bosons).

The STIs for four-point diagrams have two consequences. If the diagram
contains a four-point coupling, the resulting relation allows to
express this coupling in terms of three-point couplings. In this way,
all four-point couplings with at least one Goldstone or vector boson
can be derived; they are summarized in
Appendix~\ref{sec:other_STIs}. If the diagram does not contain a
four-point coupling, the STIs yield sum rules which imply additional
relations among three-point couplings. For instance, the Lie-algebra
structure of the vector and fermion couplings is reflected by the two
sum rules
\begin{align}
  & \isum{v_5}\, \big( g_{v_1v_2v_5} g_{v_3v_4 \bar v_5}
  + g_{v_2v_3v_5} g_{v_1v_4 \bar v_5}
  + g_{v_3v_1v_5} g_{v_2v_4 \bar v_5} \big) = 0 \,,
  \label{eq:sumrule_Jacobi} \\
  & \isum{v_3}\, g_{v_3 \bar f_1 f_2}^{\sigma} g_{v_1 v_2 \bar v_3} =
  \isum{f_3} \, \big( g_{v_1 \bar f_1 f_3}^{\sigma}
  g_{v_2 \bar f_3 f_2}^{\sigma} - g_{v_2 \bar f_1 f_3}^{\sigma}
  g_{v_1 \bar f_3 f_2}^{\sigma} \big) \,.
  \label{eq:unitarity_sumrule}
\end{align}
Eq.~\eqref{eq:sumrule_Jacobi} is simply the Jacobi identity, and
Eq.~\eqref{eq:unitarity_sumrule} relates the structure constants of
fermion and vector representations. It is interesting to note that
Eq.~\eqref{eq:unitarity_sumrule} implies the unitarity of the
Cabibbo-Kobayashi-Maskawa (CKM) matrix for an universal, diagonal
$Z$-boson coupling to fermions. The remaining sum rules provide
non-trivial constraints on the unitarization properties of the given
couplings. Here we show only the two sum rules needed for the
renormalisation of the $Z$ penguin:
\begin{align}
  & \isum{s_1} \, g_{v_1v_2 \bar s_1 } y_{s_1 \bar f_1 f_2}^{\sigma} =
  \isum{v_3} \, \tfrac{M_{v_1}^2-M_{v_2}^2}{M_{v_3}^2} \,
  g_{v_1v_2 \bar v_3 } \Big( m_{f_1} g_{v_3 \bar f_1 f_2}^{\sigma} -
  g_{v_3 \bar f_1 f_2}^{\ol\sigma} m_{f_2} \Big) \notag\\
  &\mkern+135mu + \isum{f_3} \Big( - m_{f_1} \big(
  g_{v_2 \bar f_1 f_3}^{\sigma} g_{v_1 \bar f_3 f_2}^{\sigma} +
  g_{v_1 \bar f_1 f_3}^{\sigma} g_{v_2 \bar f_3 f_2}^{\sigma} \big)
  \notag\\[-4pt]
  &\mkern+187mu - m_{f_2} \big( g_{v_2 \bar f_1 f_3}^{\ol\sigma}
  g_{v_1 \bar f_3 f_2}^{\ol\sigma} + g_{v_1 \bar f_1 f_3}^{\ol\sigma}
  g_{v_2 \bar f_3 f_2}^{\ol\sigma} \big) \notag\\[3pt]
  &\mkern+187mu + 2\, m_{f_3} \big( g_{v_2 \bar f_1 f_3}^{\ol\sigma}
  g_{v_1 \bar f_3 f_2}^{\sigma} + g_{v_1 \bar f_1 f_3}^{\ol\sigma}
  g_{v_2 \bar f_3 f_2}^{\sigma} \big) \Big) \,,
  \label{eq:gauge_mass_sumrule} \displaybreak[0] \\[5pt]
  & \isum{s_1} \, g_{v_1s_2 \bar s_1 } y_{s_1 \bar f_1 f_2}^{\sigma} =
  - \isum{v_3} \, \tfrac{1}{2 M_{v_3}^2} g_{v_1 \bar v_3 s_2} \Big(
  m_{f_1} g_{v_3 \bar f_1 f_2}^{\sigma} - g_{v_3 \bar f_1 f_2}^{\ol\sigma}
  m_{f_2} \Big) \notag\\
  &\mkern+133mu + \isum{f_3} \Big( g_{v_1 \bar f_1 f_3}^{\ol\sigma}
  y_{s_2 \bar f_3 f_2}^{\sigma} - y_{s_2 \bar f_1 f_3}^{\sigma}
  g_{v_1 \bar f_3 f_2}^{\sigma} \Big) \,.
  \label{eq:yukawa_sumrule}
\end{align}
We refer to Appendix~\ref{sec:other_STIs} for the remaining rules.

\section{Result for the $Z$ penguin}
\label{sec:general_result}

\begin{figure}[t]
  \begin{center}
    \includegraphics[width=0.32\textwidth]{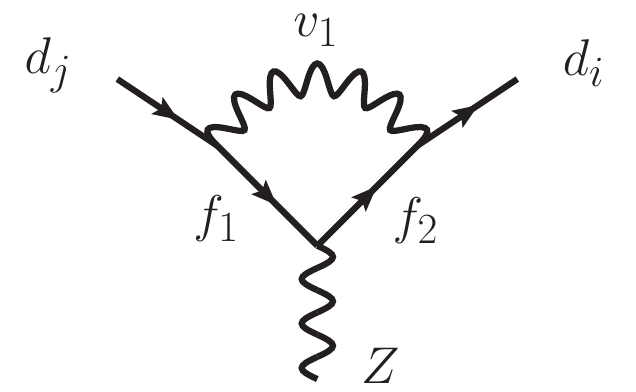}
    \includegraphics[width=0.32\textwidth]{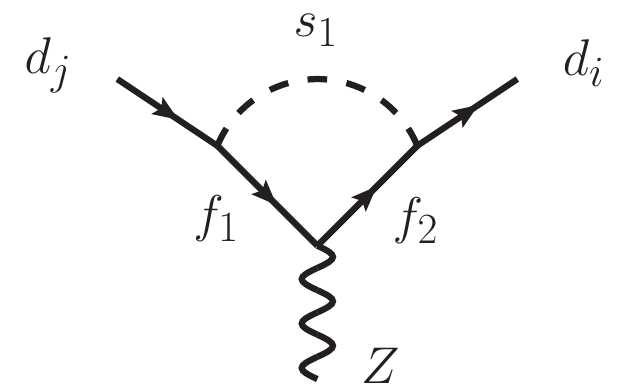}
    \includegraphics[width=0.32\textwidth]{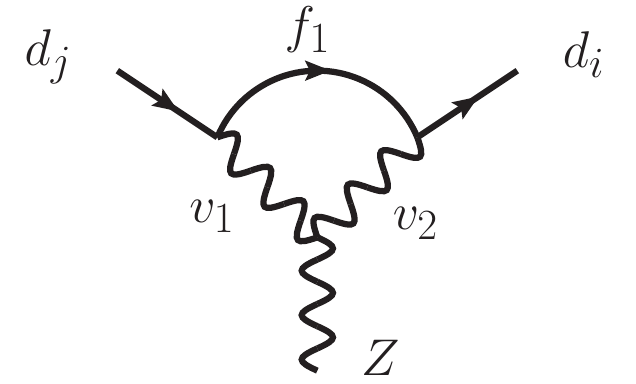}
  \end{center}
  \begin{center}
    \includegraphics[width=0.32\textwidth]{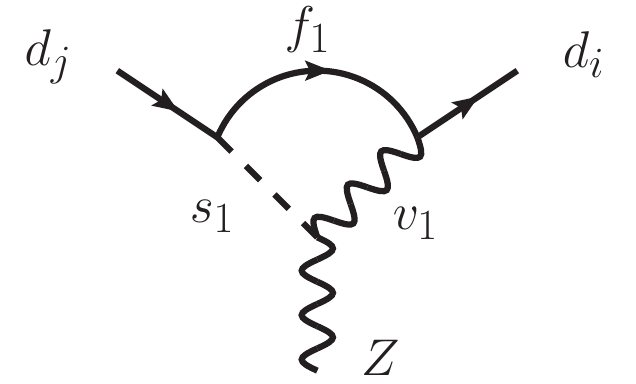}
    \includegraphics[width=0.32\textwidth]{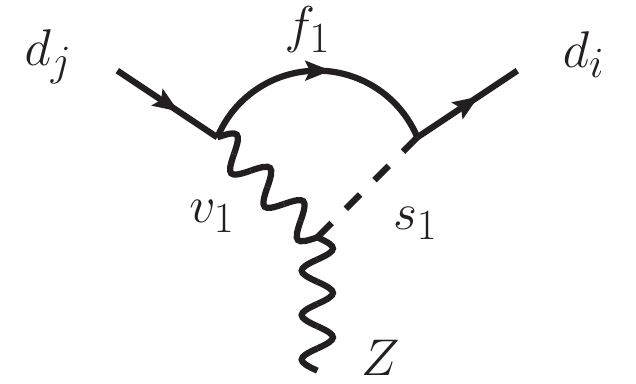}
    \includegraphics[width=0.32\textwidth]{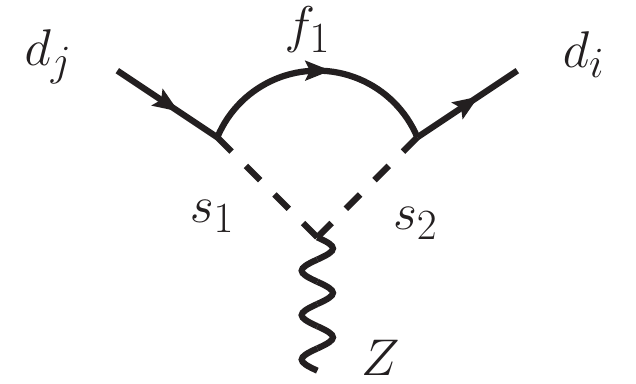}
    \caption{Representative Feynman diagrams contributing to the
      one-loop amputated $d_j - d_i - Z$ Green's function. Straight
      lines denote fermions, wavy lines denotes vectors, and dashed
      lines denote scalar particles.
        \label{fig:Zdiagrams}}
  \end{center}
\end{figure}

We present our result for the renormalised $Z$ penguin in the form of
an effective vertex. The penguin function $C^\sigma_{d_{j}d_{i}Z}$
describing the transition between two light SM fermions $d_j \to d_i$
is defined in terms of the amputated vertex function as
\begin{equation}
  \label{eq:sm_cfunction}
  \Gamma_{\mu \, \sigma}^{d_{j}d_{i}Z} = \gamma_\mu P_\sigma \times
  \frac{G_F}{\sqrt{2}} \frac{e}{\pi^2} M_Z^2 \frac{c_w}{s_w}
  C^\sigma_{d_{j}d_{i}Z} \equiv \frac{\gamma_\mu P_\sigma}{(4\pi)^2}
  \hat C^\sigma_{d_{j}d_{i}Z}\,.
\end{equation}
The function $C^\sigma_{d_{j}d_{i}Z}$, or $\hat
C^\sigma_{d_{j}d_{i}Z}$ respectively, depends on all masses and
couplings that appear; $\sigma = L,R$ stands for the chiral
projection. It is obtained by calculating the Feynman diagrams shown
in Fig.~\ref{fig:Zdiagrams}, using the Feynman rules derived from the
Lagrangian Eq.~\eqref{eq:generic_lagrangian}. In the physical meson
decay process, the momentum transfer is much smaller than any of the
internal particle masses. Accordingly, setting the external momenta
and light fermion masses to zero in the matching calculation, the
result in the 't~Hooft-Feynman gauge is
\begin{equation}\label{eq:zpenguin-result}
  \begin{split}
    &\hat C^\sigma_{d_{j}d_{i}Z} =
    \Bigg\{  
    \sum_{f_1v_1} \sum_{f_2\notin\,\rm SM} \bigg[ \kappa^\sigma_{f_1f_2v_1}
    B_0\big(m_{f_1}, m_{v_1}\big) + \kappa'^\sigma_{f_1f_2v_1} \bigg]
    + \sum_{f_1s_1} \sum_{f_2\notin\,\rm SM} \kappa^\sigma_{f_1f_2s_1}
    B_0\big(m_{f_1}, m_{s_1}\big) 
    \\ 
    &\quad + \sum_{f_1f_2v_1} \bigg[
    \tilde k^\sigma_{f_1f_2v_1} \Big(\tilde{C_0}
      \big(m_{f_1},m_{f_2},M_{v_1}\big)-\tfrac{1}{2}\Big) +
    k^\sigma_{f_1f_2v_1} C_0\big(m_{f_1},m_{f_2},M_{v_1}\big) +
    k'^{\sigma}_{f_1f_2v_1} \bigg] \\
    &\quad + \sum _{f_1v_1v_2} \bigg[
    \tilde k^\sigma_{f_1v_1v_2} \Big(\tilde{C_0}
      \big(m_{f_1},M_{v_1},M_{v_2}\big)+\tfrac{1}{2}\Big) +
    k^\sigma_{f_1v_1v_2} C_0\big(m_{f_1},M_{v_1},M_{v_2}\big) +
    k'^{\sigma}_{f_1v_1v_2} \bigg] \\
    &\quad + \sum_{f_1v_1s_1} \bigg[
    \tilde k^\sigma_{f_1v_1s_1} \Big(\tilde{C_0}
      \big(M_{s_1},m_{f_1},M_{v_1}\big)+\tfrac{1}{2}\Big) +
    k^\sigma_{f_1v_1s_1} C_0\big(M_{s_1},m_{f_1},M_{v_1}\big) \bigg] \\
    &\quad + \sum_{f_1f_2s_1} \bigg[
    \tilde k^\sigma_{f_1f_2s_1} \Big(\tilde{C_0}
      \big(M_{s_1},m_{f_1},m_{f_2}\big)-\tfrac{1}{2}\Big) +
    k^\sigma_{f_1f_2s_1} C_0\big(M_{s_1},m_{f_1},m_{f_2}\big) \bigg] \\
    &\quad + \sum_{f_1s_1s_2}
    \tilde k^\sigma_{f_1s_1s_2} \Big(\tilde{C_0}
      \big(M_{s_1},M_{s_2},m_{f_1}\big)+\tfrac{1}{2}\Big) \Bigg\} \,.
  \end{split}
\end{equation}
The functions $C_0$ and $\tilde C_0$ can be found in
Eqs.~\eqref{eq:loop-expl} and~\eqref{eq:relation}, respectively. The
couplings are contained in the $\kappa$ and $k$ factors that are
defined as follows:
\begin{equation}\label{eq:kappa-ffv}
  \begin{split}
  \kappa^\sigma_{f_1f_2v_1} &= -\frac{m_{f_1}}{m_{f_2}} \bigg(
    g_{Z \bar d_i f_2}^\sigma \bigg[ \bigg( 4 - \frac{m_{f_1}^2}{M_{v_1}^2}
    \bigg) g_{v_1 \bar f_2 f_1}^{\ol\sigma} + \frac{m_{f_1} m_{f_2}}{M_{v_1}^2}
    g_{v_1 \bar f_2 f_1}^{\sigma} \bigg] g_{\bar v_1 \bar f_1 d_j}^{\sigma} \\
    &\mkern+60mu +
    g_{\bar v_1 \bar d_i f_1}^{\sigma} \bigg[ \bigg( 4 - \frac{m_{f_1}^2}{M_{v_1}^2}
    \bigg) g_{v_1 \bar f_1 f_2}^{\ol\sigma} + \frac{m_{f_1} m_{f_2}}{M_{v_1}^2}
    g_{v_1 \bar f_1 f_2}^{\sigma} \bigg] g_{Z \bar f_2 d_j}^\sigma \bigg)
    \,,\\
  \kappa'^\sigma_{f_1f_2v_1} &= \frac{2m_{f_1}}{m_{f_2}} \Big(
    g_{Z \bar d_i f_2}^{\sigma} g_{v_1 \bar f_2 f_1}^{\ol\sigma}
    g_{\bar v_1 \bar f_1 d_j}^{\sigma} + g_{v_1 \bar d_i f_1}^{\sigma}
    g_{\bar v_1 \bar f_1 f_2}^{\ol\sigma} g_{Z \bar f_2 d_j}^{\sigma} \Big)
    \,,\\
  \end{split}
\end{equation}
\begin{equation}\label{eq:kappa-ffs}
  \begin{split}
  \kappa^\sigma_{f_1f_2s_1} &= \frac{m_{f_1}}{m_{f_2}} \Big(
    g_{Z \bar d_i f_2}^\sigma y_{s_1 \bar f_2 f_1}^{\sigma}
    y_{\bar s_1 \bar f_1 d_j}^{\sigma} + y_{s_1 \bar d_i f_1}^{\ol\sigma}
    y_{\bar s_1 \bar f_1 f_2}^{\ol\sigma} g_{Z \bar f_2 d_j}^\sigma \Big)
    \,,\\
  \end{split}
\end{equation}
\begin{equation}\label{eq:k-ffv}
  \begin{split}
  \tilde k^\sigma_{f_1f_2v_1} &= \bigg(
    g_{Z\bar{f}_2f_1}^{\sigma} + \frac{m_{f_1} m_{f_2}}{2 M_{v_1}^2}
    g_{Z\bar{f}_2f_1}^{\ol\sigma} \bigg) \, g_{\bar v_1 \bar d_i f_2}^{\sigma}
    g_{v_1 \bar f_1 d_j}^{\sigma} \,,\\
  k^\sigma_{f_1f_2v_1} &= -\frac{m_{f_1} m_{f_2}}{M_{v_1}^2} \,
  \big( m_{f_1} m_{f_2} g_{Z\bar{f}_2f_1}^{\sigma}
    + 2 M_{v_1}^2 g_{Z\bar{f}_2f_1}^{\ol\sigma} \big) \,
  g_{\bar v_1 \bar d_i f_2}^{\sigma} g_{v_1 \bar f_1 d_j}^{\sigma} \,,\\
  k'^{\sigma}_{f_1f_2v_1} &= -g_{Z\bar{f}_2f_1}^{\sigma}
  g_{\bar v_1 \bar d_i f_2}^{\sigma} g_{v_1 \bar f_1 d_j}^{\sigma} 
   + \frac{1}{2}
    \big( g_{Z \bar d_i d_i}^\sigma g_{v_1 \bar d_i
      f_1}^{\sigma} g_{\bar v_1 \bar f_1 d_j}^\sigma + g_{\bar v_1
      \bar d_i f_1}^\sigma  g_{v_1 \bar f_1 d_j}^{\sigma} g_{Z \bar
      d_j d_j}^\sigma \big) \delta_{f_1 f_2}
    \,,\\
  \end{split}
\end{equation}
\begin{equation}\label{eq:k-fvv}
  \begin{split}
  \tilde k^\sigma_{f_1v_1v_2} &= - \bigg(3 + \frac{m_{f_1}^2 (M_{v_1}^2+M_{v_2}^2-M_Z^2)}
    {4 M_{v_1}^2 M_{v_2}^2}\bigg) \, g_{Z v_1 \bar v_2}
    g_{\bar v_1 \bar d_i f_1}^{\sigma} g_{v_2 \bar{f}_1
      d_{j}}^{\sigma} \\
  & \quad - \frac{1}{2} \bigg( 1 + \frac{m_{f_1}^2}{2M_{v_1}^2}
    \bigg) \big( g_{Z \bar d_i d_i}^\sigma 
    g_{v_1 \bar d_i f_1}^{\sigma} g_{\bar v_1 \bar f_1 d_j}^\sigma 
    + g_{v_1 \bar d_i f_1}^\sigma  
    g_{\bar v_1 \bar f_1 d_j}^{\sigma} g_{Z \bar d_j d_j}^\sigma \big)
    \delta_{v_1 v_2} \,,\\
  k^\sigma_{f_1v_1v_2} &= \frac{m_{f_1}^2
    \left(M_{v_1}^4+M_{v_2}^4-M_Z^2(M_{v_1}^2+M_{v_2}^2)\right)
    }{M_{v_1}^2 M_{v_2}^2} \, g_{Z v_1 \bar v_2}
    g_{\bar v_1 \bar d_i f_1}^{\sigma} g_{v_2 \bar f_1 d_j}^{\sigma}
    \,,\\
  k'^{\sigma}_{f_1v_1v_2} &= 2 \, g_{Z v_1 \bar{v}_2}
    g_{\bar v_1 \bar d_i f_1}^{\sigma} g_{v_2 \bar f_1 d_j}^{\sigma} 
    \,,\\
  \end{split}
\end{equation}
\begin{equation}\label{eq:k-fvs}
  \begin{split}
  \tilde k^\sigma_{f_1v_1s_1} &= - \frac{k^\sigma_{f_1v_1s_1}}{4 M_{v_1}^2} \,,
  \qquad k^\sigma_{f_1v_1s_1} = m_{f_1}\left(
    g_{Z \bar v_1 s_1} y_{\bar s_1 \bar d_i f_1}^{\ol\sigma}
    g_{v_1 \bar f_1 d_j}^{\sigma} + g_{Z v_1 \bar s_1} y_{s_1 \bar f_1 d_j}^{\sigma}
    g_{\bar v_1 \bar d_i f_1}^{\sigma} \right) \,,\\
  \end{split}
\end{equation}
\begin{equation}\label{eq:k-ffs}
  \begin{split}
  \tilde k^\sigma_{f_1f_2s_1} &= \frac{1}{2}\, g_{Z \bar f_2 f_1}^{\ol\sigma}
  y_{\bar s_1 \bar d_i f_2}^{\ol\sigma} y_{s_1 \bar f_1 d_j}^{\sigma} \, ,
  \qquad k^\sigma_{f_1f_2s_1} = - m_{f_1} m_{f_2} g_{Z \bar f_2 f_1}^{\sigma}
  y_{\bar s_1 \bar d_i f_2}^{\ol\sigma} y_{s_1 \bar f_1 d_j}^{\sigma}
  \,, \\ 
  \end{split}
\end{equation}
\begin{equation}\label{eq:k-fss}
  \begin{split}
  \tilde k^\sigma_{f_1s_1s_2} & = \frac{1}{2}\, g_{Z \bar s_1 s_2}
  y_{\bar s_2 \bar d_i f_1}^{\ol\sigma} y_{s_1 \bar f_1 d_j}^{\sigma}
  -\frac{1}{4} \Big( g_{Z \bar d_i
    d_i}^\sigma y_{\bar s_1 \bar d_i f_1}^{\ol\sigma} 
    y_{s_1 \bar f_1 d_j}^\sigma + y_{\bar s_1 \bar d_i f_1}^{\ol\sigma}
    y_{s_1 \bar f_1 d_j}^{\sigma} g_{Z \bar d_j d_j}^\sigma \Big)
    \delta_{s_1 s_2} \,.
  \end{split}
\end{equation}
The sums in~\eqref{eq:zpenguin-result} run over all components of the
fields (e.g., explicitly over $W^+$ and $W^-$ in the SM). Note that
the scalar-index sums in~\eqref{eq:zpenguin-result} run only over
physical fields; the contributions of would-be Goldstone bosons have
been accounted for via the replacement of the Goldstone couplings in
terms of couplings to physical particles, using
Eq.~\eqref{eq:STI_3pt_GB_replacements}. The contribution of the $Z$
penguin to processes like rare $B$ and $K$ decays is not separately
gauge-invariant without including the box contributions. We present
the explicit generic results for completeness in
Appendix~\ref{sec:app_box}.

Our result~\eqref{eq:sm_cfunction} consists of several
contributions. In general, we can write ($i \neq j$)
\begin{equation} \label{eq:penguin_divergence_unsimplified}
  \begin{split}
    \Gamma_{\mu \, \sigma}^{d_j d_i Z} 
    = \Gamma_{\mu \, \sigma}^{d_j d_i Z, (1)}
    + \Big[ \delta Z_{Z \bar d_i d_j}^{g, \sigma} +
    \tfrac12 \isum{f_1} \Big( g_{Z \bar d_i f_1}^{\sigma} \delta
    Z_{\bar f_1 d_j}^{\psi,\sigma} + \delta Z_{\bar d_i f_1}^{\psi,\sigma\, \dagger}
    g_{Z \bar f_1 d_j}^{\sigma} \Big) + \tfrac12 \isum{v_1} \delta
    Z_{Z v_1}^{V} g_{\bar v_1 \bar d_i d_j}^{\sigma} \Big] P_\sigma \,.
  \end{split}
\end{equation}
Here $\Gamma_{\mu \, \sigma}^{Z d_j d_i, (1)}$ denotes the sum of all
contributing one-loop diagrams, and the rest are the vertex, fermion
field, and gauge-boson field renormalisation constants,
respectively. Note that in this work we assume the absence of
tree-level FCNC transitions between light quarks, i.e. $g_{\bar v_1
  \bar d_i d_j}^{\sigma} = 0$. In this case, we can use the
STIs~\eqref{eq:unitarity_sumrule}, \eqref{eq:gauge_mass_sumrule}
and~\eqref{eq:yukawa_sumrule} to show that the fermion field
renormalisation constants are sufficient to cancel all divergences
in~\eqref{eq:penguin_divergence_unsimplified} (in other words, $\delta
Z_{Z \bar d_i d_j}^{g, \sigma}$ can be chosen to vanish). This is done
explicitly in Sec.~\ref{sec:specific_models}.

\begin{figure}[t]
  \begin{center}
    \includegraphics[width=0.35\textwidth]{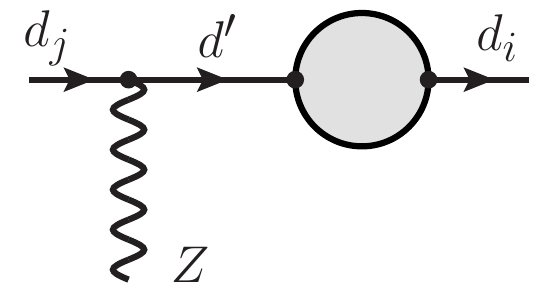}
      \caption{One-heavy-particle reducible contribution to the $Z$
        penguin. 
        \label{fig:1PR}}
  \end{center}
\end{figure}

The two-point functions in the full theory lead to mixing among the
fermions beyond tree level. We perform an off-diagonal field
renormalisation for all fermions fields, with finite terms chosen to
restore a diagonal and canonically normalised kinetic and mass
term. As one consequence, diagrams with a reducible heavy-fermion line
(of the form shown in Figure~\ref{fig:1PR}) are exactly cancelled by
corresponding counterterm diagrams.

The one-loop corrections to the renormalised fermion two-point
function can be written as
\begin{equation}\label{eq:two-point}
    \delta\Gamma_{\sigma}^{f_i f_j}
    = \Big[ \big( \Sigma_{V,\bar f_j f_i}^{\psi,\sigma} +
    \delta Z_{H,\bar f_j f_i}^{\psi,\sigma} \big) \slashed{p} +
    \Sigma_{S, \bar f_j f_i}^{\psi,\sigma} 
    - \tfrac12 (m_{f_j} \delta Z_{\bar f_j f_i}^{\psi,\sigma} +
    \delta Z_{\bar f_j f_i}^{\psi,\ol\sigma \, \dagger} m_{f_i}) - \delta_{ji}
    \delta m_{f_i} \Big] P_\sigma \,.
\end{equation}
The finite parts of the field renormalisation constants for $j\neq i$
are obtained from the requirement that the off-diagonal parts of the
self energy~\eqref{eq:two-point} vanish. They are given by
\begin{equation}
  \delta Z_{\bar f_jf_i}^{\psi,\sigma}\big|_{\rm fin} =
  \frac{2}{m_{f_j}^2 - m_{f_i}^2} \Big( m_{f_i}^2 \Sigma_{V,\bar
    f_jf_i}^{\psi,\sigma} +  m_{f_j} m_{f_i} \Sigma_{V,\bar
    f_jf_i}^{\psi,\ol\sigma} + m_{f_j} \Sigma_{S,\bar
    f_jf_i}^{\psi,\sigma} +  m_{f_i} \Sigma_{S,\bar
    f_jf_i}^{\psi,\ol\sigma} \Big)\big|_{\rm fin} \,.
\end{equation}
They enter our generic result after having been expanded in small mass
ratios. The diagonal field renormalisation constants are not necessary
in our case. If we had allowed for the presence of tree-level $d_j \to
d_i$ transitions, one would have to fix these constants by
renormalisation conditions. Note that, in this case, the one-loop
result would correspond to a tiny correction of an already suppressed
tree-level amplitude. These consideration can be important, however,
in an analogous treatment of charged current
couplings~\cite{Gambino:1998ec}.

We will show in the next section that all divergences in our generic
result completely cancel against the divergent terms in the field
rotation, without introducing additional counterterms. In order to
make this cancellation manifest, it is necessary to use the
consequences of tree-level perturbative unitarity derived in
Sec.~\ref{sec:generic_lagrangian} in form of the STIs. This step is
independent of the specification of a model and can be applied without
detailed knowledge about the sector responsible for the spontaneous
symmetry breaking.

\section{Renormalised Results}
\label{sec:specific_models}

The result~\eqref{eq:zpenguin-result} is valid for any number of heavy
new scalars, vectors, and Dirac fermions. It is, however, not very
suitable for numerical evaluation: the functions $\tilde{C_0}$ in
Eq.~\eqref{eq:zpenguin-SM-gen} contain divergent terms
(cf. Appendix~\ref{sec:loopfunctions}). While the STI (via relations
between the couplings appearing in the
coefficients~\eqref{eq:kappa-ffv}~-~\eqref{eq:k-fss}) ensure that the
result is finite, the cancellation of the divergences is not
manifest. In this section we will derive manifestly finite versions of
our result that, in addition, depend on a minimal number of physical
couplings.

In order to make contact to phenomenological applications, we remind
the reader that the loop functions appearing in rare $K$ and $B$ meson
decays are given by the following combinations~\cite{Buchalla:1995vs}
\begin{equation}
\begin{split}
X^{\sigma}(x) = C_{sdZ}^{\sigma}(x) - 4 B_{sd\nu\nu}^{\sigma L}(x) \,, \quad
Y^{\sigma\sigma'}(x) = C_{sdZ}^{\sigma}(x) - B(x)_{sd\ell\ell}^{\sigma \sigma'} \,,
\end{split}
\end{equation}
with $C_{sdZ}^{\sigma}$ and $B_{sd\ell_k\ell_k}^{\sigma \sigma'}$
given in Eq.~\eqref{eq:sm_cfunction} and
Eq.~\eqref{eq:generic_box_coefficients}, respectively.

\subsection{SM Fermions and charged scalars and vectors}

We first consider the simplified case of theories with charged heavy
scalars and vectors -- in other words, we drop all coefficients in
Eq.~\eqref{eq:zpenguin-result} that involve couplings to heavy new
fermions. We then simplify the remaining expression by repeatedly
eliminating physical coupling constants via application of the
STI. The result obtained in this way is not only manifestly finite,
but also depends on fewer physical coupling constants than the
original expression.  This procedure is not unique. While the STI
ensure that one will finally arrive at a finite result, there is some
freedom of which couplings to eliminate and which ones to retain.

For instance, continuing with our simplified case, we can first
eliminate the right-handed $Z$-fermion coupling by solving the
``Yukawa sum rule'' Eq.~\eqref{eq:yukawa_sumrule} for $g_{Z\bar q
q}^R$, and solving the ``gauge boson mass sum rule''
Eq.~\eqref{eq:gauge_mass_sumrule} for $g_{Z\bar t t}^R$. We then apply
the ``unitarity sum rule'', Eq.~\eqref{eq:unitarity_sumrule}, together
with the universality of the $Z$ coupling to fermions, to eliminate
all dependence on couplings involving light quarks (this step is a
generalization of the GIM mechanism). We can then solve
Eq.~\eqref{eq:unitarity_sumrule} for $g_{Z\bar t t}^L$. Applying the
resulting relations to Eqs.~\eqref{eq:kappa-ffv}-\eqref{eq:k-fss}
eliminates all divergences, and we obtain the manifestly finite
expression
\begin{equation}\label{eq:SV-explicit}
  \begin{split}
    \hat{C}^{L}_{d_jd_iZ} = & \sum_{s_1 s_2} f_S(m_t,M_{s_1},M_{s_2})
y_{s_2^+ \bar{t} d_j}^L \left(\delta_{s_1 s_2} y_{s_2^- \bar{d_i} t}^R \left(g_{Z \bar{d_j} d_j}^L -g_{Z \bar{t} t}^L\right) + g_{Z s_1^+ s_2^-} y_{s_1^- \bar{d_i} t}^R\right) \\ 
    & + \sum_{v_1 v_2} f_V(m_t,M_{v_1},M_{v_2}) g_{Z v_2^+ v_1^-} g_{v_1^+ \bar{t} d_j}^L g_{v_2^- \bar{d_i} t}^L \\
    & + \sum_{s_1 v_1} f_{VS}(m_t,M_{s_1},M_{v_1}) y_{s_1^+ \bar{t} d_j}^L g_{v_1^- \bar{d_i} t}^L g_{Z v_1^+ s_1^-} \\
    & + \sum_{s_1 v_1} f_{VS'}(m_t,M_{s_1},M_{v_1}) y_{s_1^- \bar{d_i} t}^R g_{v_1^+ \bar{t} d_j}^L g_{Z v_1^- s_1^+} \,.
  \end{split}
\end{equation}
The loop functions are given by
\begin{equation}
  \begin{split}
    f_{V}(m_i,m_j,m_k) &= m_i^2 C_0\left(m_i,m_k,m_k\right)-\frac{m_i^2 \left(m_j^2+m_k^2-M_Z^2\right)}{4 m_j^2 m_k^2} \\
    & + \frac{m_i^2 \left(-3 m_j^2+m_k^2-M_Z^2\right)+4 m_k^2\left(m_j^2-m_k^2+M_Z^2\right)}{4 m_j^2 m_k^2} m_i^2 C_0\left(m_i,m_i,m_k\right) \\
    & + \frac{-M_Z^2 \left(3 m_j^2+4 m_k^2\right)-13 m_j^2 m_k^2+3 m_j^4+4 m_k^4}{4 m_j^2 m_k^2} m_i^2 C_0\left(m_i,m_j,m_k\right)\,, \\
    f_{S}(m_i,m_j,m_k) &= \frac{1}{2} m_i^2 C_0\left(m_i,m_i,m_k\right)-\frac{1}{2} m_j^2 C_0\left(m_i,m_j,m_k\right)-\frac{1}{2}\,, \\
    f_{VS}(m_i,m_j,m_k) &= m_i \left(1-\frac{m_j^2}{4 m_k^2}\right)
    C_0\left(m_i,m_j,m_k\right) \\ & +\left(\frac{m_i^3}{4 m_k^2}-m_i\right)
    C_0\left(m_i,m_i,m_k\right)-\frac{m_i}{4 m_k^2}\,, \\ 
    f_{VS'}(m_i,m_j,m_k) &= \frac{m_i^3 C_0\left(m_i,m_i,m_j\right)}{4 m_k^2}+\frac{3}{4} m_i C_0\left(m_i,m_j,m_k\right)-\frac{m_i}{4 m_k^2}\,.
  \end{split}
\end{equation}
The function $C_0$ can be found in Eq.~\eqref{eq:loop-expl}. 

A comment is in order: The sums over charged vectors $v_i$ and charged
scalars $s_i$ in Eq.~\eqref{eq:SV-explicit} run over the particle
types, but not over the different charges: for SM fermion content the
charge of the internal vectors and scalars is fixed by the external
fermions (e.g., for a strange to down transition, there are only
up-type quarks in the loop, so the internal vectors and scalars must
have negative charge $-e$).

\subsubsection{Standard Model}\label{sec:SM}

We illustrate the general procedure described above by considering the
$s \to d$ transition in the SM. Here, the only relevant heavy degrees
of freedom are the top quark, and the gauge bosons $Z$ and
$W^\pm$. Starting with the general result~\eqref{eq:zpenguin-result},
we need to keep only the terms
\begin{equation}
  \label{eq:zpenguin-SM-gen}
  \begin{split}
    \hat{C}^{L,\text{SM}}_{sdZ} =
    & \sum_{f_1f_2v_1} \bigg[
    \tilde k^\sigma_{f_1f_2v_1} \Big(\tilde{C_0}
      \big(m_{f_1},m_{f_2},M_{v_1}\big)-\tfrac{1}{2}\Big) +
    k^\sigma_{f_1f_2v_1} C_0\big(m_{f_1},m_{f_2},M_{v_1}\big) +
    k'^{\sigma}_{f_1f_2v_1} \bigg] \\
    + & \sum _{f_1v_1v_2} \bigg[
    \tilde k^\sigma_{f_1v_1v_2} \Big(\tilde{C_0}
      \big(m_{f_1},M_{v_1},M_{v_2}\big)+\tfrac{1}{2}\Big) +
    k^\sigma_{f_1v_1v_2} C_0\big(m_{f_1},M_{v_1},M_{v_2}\big) +
    k'^{\sigma}_{f_1v_1v_2} \bigg] \,
  \end{split}
\end{equation}
Next, we insert the SM couplings of $W$ bosons
\begin{equation}
  \label{eq:wqq_smcoupling}
  g^L_{W^+ \bar u_j d_k} = \frac{e}{s_w \sqrt{2}} V_{jk}
  \,, \qquad
  g^L_{W^+ \bar \nu_j \ell_k} = \frac{e}{s_w \sqrt{2}} \delta_{jk}
  \,, \qquad
  g_{Z W^+ W^-} = \frac{e}{t_w}
\end{equation}
and $Z$ bosons
\begin{equation}
  \label{eq:zqq_smcoupling}
  g^L_{Z \bar f_j f_k} = \frac{2 e}{s_{2 w}}
  \left( T_3^f - Q_f s_w^2 \right) \delta _{jk}
  \,, \qquad
  g^R_{Z \bar f_j f_k} = - \frac{2 e}{s_{2 w}}
  Q_f s_w^2 \delta _{jk} \, .
\end{equation}
Here $T_3^f=\pm 1/2$ is the third component of the weak isospin of the
fermion $f$ and $Q_f$ is its electric charge in units of the positron
charge $e$. The sine of the weak mixing angle is denoted by $s_w =
\sin(\theta_w)$. In addition we defined $s_{2w} = \sin(2\theta_w)$ and
$t_{w} = \tan(\theta_w)$. The CKM matrix elements are denoted by
$V_{jk}$.

The $Z$-boson coupling to fermions in the SM is universal and
diagonal; hence, the sum rule~\eqref{eq:unitarity_sumrule} leads to the
unitarity of the CKM matrix. Defining $\lambda_q^{(d'd)} = V_{qd'}
V_{qd}^*$, this allows us to eliminate $\lambda_{u}^{sd} = -
\lambda_{c}^{sd} - \lambda_{t}^{sd}$, and thus the coefficient
$\lambda_{t}^{sd}$ multiplies the difference of the diagrams with
massive top and massless up quark. This leads to a partial
cancellation of UV divergences -- the well-known GIM
(Glashow-Iliopoulos-Maiani) mechanism.

After this manipulation, the result still has a left-over divergence,
proportional to
\begin{equation}
  \frac{1}{M_W^2}-\frac{1}{c_w^2 M_Z^2} \, .
\end{equation}
Evaluating the sum rule~\eqref{eq:gauge_mass_sumrule} for the choice
$\sigma=L$, $v_1=Z$, $v_2=W^+$, $f_1=t$, $f_2=b$, and inserting the SM
couplings, leads immediately to the relation
\begin{equation}\label{eq:unitarity:SM}
  M_W^2 = c_w^2 M_Z^2 \, .
\end{equation}
Thus, the divergent term vanishes. The finite term exactly reproduces
the result of Inami and
Lim~\cite{Inami:1980fz}
\begin{equation} \label{eq:sm_zpenguin}
   C^{L\,(\rm SM)}_{d'dZ} = \lambda_t^{(d'd)} \, C(x_t) \,,\quad
   C(x) = \frac{x}{8} \bigg[ \frac{x - 6}{x - 1} +
     \frac{3 x + 2}{(x - 1)^2} \log(x) \bigg] \,,
\end{equation}
for the top-quark contribution to the $d' \to d Z$ vertex. The ratio
of the quark and $W$-boson masses squared is denoted by
$x_q=m_q^2/M_W^2$.

Of course, the same result is obtained in a much simpler way by
directly using the result~\eqref{eq:SV-explicit}. Both the GIM
mechanism and ``gauge boson mass relation'' \eqref{eq:unitarity:SM}
are already built in. Moreover, it suffices to specify the reduced set
of SM couplings~\eqref{eq:wqq_smcoupling} -- the $Z$-fermion couplings
\eqref{eq:zqq_smcoupling} are then fixed via the STI.

\subsubsection{2HDM}

As an example with additional charged scalars we consider the
contribution of charged Higgs bosons in a two Higgs doublet model
(2HDM)~\cite{Gunion:1989we}. Using that the couplings involving both
charged gauge and Higgs bosons vanish, $g_{Z W_j^+ h_i^-} = g_{Z W_j^-
  h_i^+} = 0$, we see that only the first line in
Eq.~\eqref{eq:SV-explicit} contributes, with a prefactor given by
\begin{equation} 
 y_{h_j^+ \bar{t} d}^L \left(\delta_{i j} y_{h_j^- \bar{s} t}^R \left(g_{Z \bar{d} d}^L -g_{Z \bar{t} t}^L\right)
+g_{Z h_i^+ h_j^-} y_{h_i^- \bar{s} t}^R\right)\,.
\end{equation}
Specializing to the case of one charged Higgs, we need the following
additional Feynman rules~\cite{Gunion:1989we}
\begin{equation}
g_{Z h^- h^+} = - e \frac{c_{2w}}{2 s_w c_w} \,, \quad
y_{h^+\bar{t}d_i}^L = \big(y_{h^-\bar{d_i}t}^R\big)^* =
\frac{m_t}{t_{\beta}} \frac{V_{td} e}{\sqrt{2} s_w M_W}\,, \quad
y_{h^+\bar{t}d_i}^R = \big(y_{h^-\bar{d_i}t}^L\big)^* = {\mathcal
  O}(m_{d_i}) \,,
\end{equation}
with $t_\beta = \tan\beta$. We then find 
\begin{equation}
C(x_t, y_t) = - \frac{V_{td} V_{ts}^*}{8 t_\beta^2} \, x_t \,
f_S(m_t,M_h,M_h)
\end{equation}
where
\begin{equation}
f_S(m_t,M_h,M_h) = \frac{y_t}{1-y_t} + \frac{y_t \log y_t}{(y_t-1)^2}
\end{equation}
and we defined $x=m_t^2/M_W^2$, $y=m_t^2/M_h^2$. This reproduces the
function
\begin{equation}
C_H(x_t,y_t) = -\frac{1}{8} |Y|^2 x_t y_t \bigg[ \frac{1}{1-y_t} +
\frac{\log y_t}{(y_t-1)^2} \bigg]
\end{equation}
from~\cite{Buchalla:1990fu}, where $Y = v_1/v_2 = \cot\beta$.

\subsection{Arbitrary charged Fermions, Scalars, and Vectors}

We now consider the general case of theories with arbitrary numbers of
heavy charged fermions, scalars and vectors, and simplify the general
result Eq.~\eqref{eq:zpenguin-result} by repeatedly eliminating
physical coupling constants via application of the STI. 

In analogy to our procedure described in the previous section, we can
first eliminate the diagonal right-handed $Z$-fermion coupling by
solving the ``Yukawa sum rule'' Eq.~\eqref{eq:yukawa_sumrule} for
$g_{Z\bar f f}^R$, thus eliminating this coupling in conjunction with
couplings to heavy scalars. Next, we solve the ``gauge boson mass sum
rule'' Eq.~\eqref{eq:gauge_mass_sumrule} for $g_{Z\bar f f}^R$ and use
the resulting relation to eliminate all diagonal right-handed $Z$
couplings to fermions in conjunction with vector couplings. We then
repeatedly apply the ``unitarity sum rule'',
Eq.~\eqref{eq:unitarity_sumrule}, to eliminate all couplings of
fermions to the charged vectors and all couplings of the $Z$ boson to
left-handed fermions, looping over all heavy fermions. Together with
our assumption of universality and diagonality of the $Z$-boson
couplings to down-type SM quarks, this eliminates all divergences.
Note that the ``unitarity sum rule'',
Eq.~\eqref{eq:unitarity_sumrule}, leads to a ``generalized GIM
mechanism'', effectively eliminating some of the couplings of one
(arbitrarily chosen) heavy fermions in the loop (denoted below by the
subscript $f'$). The resulting, manifestly finite expression for the
$Z$ penguin for an arbitrary number of charged fermions, scalars, and
vectors is then
\begin{equation}\label{eq:res:general}
\begin{split}
     \hat C_{d_jd_iZ}^L = & \quad \sum_{f_1f_2v_1}
     g_{Z\bar{f}_2f_1}^L g_{v_1 \bar{f}_1d_j}^L g_{\bar
     v_1 \bar{d_i}f_2}^L F_V\left(m_{f_1},m_{f_2},M_{v_1}\right) \\ &
     + \sum_{f_1f_2v_1} g_{Z\bar{f}_2f_1}^R
     g_{v_1 \bar{f}_1d_j}^L g_{\bar v_1 \bar{d_i}f_2}^L
     F_{V'}\left(m_{f_1},m_{f_2},M_{v_1}\right) \\ &
     + \sum_{f_1v_1v_2} g_{Zv_2 \bar v_1 }
     g_{v_1 \bar{f}_1d_j}^L g_{\bar v_2 \bar{d_i}f_1}^L
     F_{V''}\left(m_{f'},m_{f_1},M_{v_1},M_{v_2} \right) \\ &
     + \sum_{f_1f_2s_1} g_{Z\bar{f}_2f_1}^L y_{s_1 \bar{f}_1d_j}^L
     y_{\bar s_1 \bar{d_i}f_2}^R
     F_S\left(m_{f_1},m_{f_2},M_{s_1}\right) \\ &
     + \sum_{f_1s_1s_2}\left( g_{Zs_2 \bar s_1 } + \delta_{s_1s_2}
     g_{Z\bar{d}_jd_j}^L \right) y_{s_1 \bar{f}_1d_j}^L y_{\bar
     s_2 \bar{d_i}f_1}^R F_{S'}\left(m_{f_1},M_{s_1},M_{s_2}\right) \\
     & + \sum_{f_1f_2s_1} g_{Z\bar{f}_2f_1}^R
     y_{s_1 \bar{f}_1d_j}^L y_{\bar s_1 \bar{d_i}f_2}^R
     F_{S''}\left(m_{f_1},m_{f_2},M_{s_1}\right) \\ &
     + \sum_{f_1s_1v_1} g_{Zv_1 \bar s_1 } y_{s_1 \bar{f}_1d_j}^L
     g_{\bar v_1 \bar{d_i}f_1}^L
     F_{\text{SV}}\left(m_{f_1},M_{s_1},M_{v_1}\right) \\ &
     + \sum_{f_1s_1v_1} g_{Z\bar v_1 s_1 } g_{v_1 \bar{f}_1d_j}^L
     y_{\bar s_1 \bar{d_i}f_1}^R
     F_{\text{SV}'}\left(m_{f_1},M_{s_1},M_{v_1}\right) \,,
\end{split}
\end{equation}
The respective loop functions can be written
\begin{equation}
\begin{split}
   & F_V\left(m_i,m_j,m_k\right) = \tilde{C}_{\delta }\left(m_i,m_j,m_k\right)+\frac{m_j^2}{m_k^2}\left(m_j^2
   C_0\left(m_j,m_j,m_k\right)-m_i^2 C_0\left(m_i,m_j,m_k\right)\right)\,,  \\
   & F_{V'}\left(m_i,m_j,m_k\right) = \frac{m_i m_j}{2 m_k^2}  \left(\tilde{C}_{\delta }\left(m_i,m_j,m_k\right)-4 m_k^2 C_{\delta} \left(m_i,m_j,m_k\right)\right)\,, \\
   & F_{V''}(m_i,m_j,m_k,m_l) = \tilde{C}_{\Delta}(m_i,m_j,m_k,m_l) + f_{V''}(m_i,m_k,m_l) - f_{V''}(m_j,m_k,m_l)\,, \\
   & F_S\left(m_i,m_j,m_k\right) = -m_i m_j C_0\left(m_i,m_j,m_k\right)\,, \\
   & F_{S'}\left(m_i,m_j,m_k\right) = - \frac{1}{2} \left(\tilde{C}_{\delta }\left(m_j,m_i,m_k\right)+1\right)\,, \\
   & F_{S''}\left(m_i,m_j,m_k\right) = \frac{1}{2} \tilde{C}_{\delta }\left(m_i,m_j,m_k\right)\,, \\
   & F_{\text{SV}}\left(m_i,m_j,m_k\right) = \frac{m_i}{4 m_k^2} \left(4 m_k^2 C_{\delta
   }\left(m_j,m_i,m_k\right) - \tilde{C}_{\delta }\left(m_j,m_i,m_k\right)-1\right)\,, \\
   & F_{\text{SV}'}\left(m_i,m_j,m_k\right) = \frac{m_i}{4 m_k^2} \left(4 m_k^2
   C_0\left(m_i,m_j,m_k\right) -\tilde{C}_{\delta }\left(m_k,m_i,m_j\right)-1\right)\,, \\
    \end{split}
\end{equation}
where
\begin{equation}
    \begin{split}
   \frac{1}{m_i^2} f_{V''}(m_i,m_j,m_k) &=  \frac{m_j^2+m_k^2-M_Z^2}{4 m_j^2 m_k^2}  
      \big(1 + \tilde{C}_{\delta} (m_j,m_i,m_k)\big) \\
     & + \left( \frac{m_i^2}{m_k^2}+\frac{m_k^2-M_Z^2}{m_j^2}+1 \right) C_0(m_i,m_i,m_k) \\
     & - \frac{m_j^4+m_k^4-M_Z^2 \left(m_j^2+m_k^2\right)}{m_j^2 m_k^2} C_0(m_i,m_j,m_k)
    \end{split}
\end{equation}
and
\begin{equation}
      \begin{split}
     \tilde{C}_\Delta\left(m_i,m_j,m_k,m_l\right) &= 
    3 \left( \tilde{C}_0\left(m_i,m_k,m_l\right) - \tilde{C}_0\left(m_j,m_k,m_l\right) \right) \\ & \quad -\tilde{C}_0\left(m_i,m_i,m_l\right) + \tilde{C}_0\left(m_j,m_j,m_l\right) \\
   \tilde{C}_\delta (m_i,m_j,m_k)&= \tilde{C}_0\left(m_i,m_j,m_k\right)-\tilde{C}_0\left(m_j,m_j,m_k\right) \\
  C_\delta (m_i,m_j,m_k) &= C_0\left(m_i,m_j,m_k\right)-C_0\left(m_j,m_j,m_k\right)\,.
      \end{split}
\end{equation}
These loop functions are manifestly finite; it is interesting to note
that one could perform the whole calculation without regulator in four
space-time dimensions if one applies the sum rules first. Note also
that several of the loop functions vanish if the fermion masses are
equal; namely, we have
\begin{equation}F_V\left(m,m,M\right) =
F_{V'}\left(m,m,M\right) = F_{V''}\left(m,m,M\right) =
F_{S''}\left(m,m,M\right) = 0\,.
\end{equation}
As a consequence, the result depends only the off-diagonal
right-handed couplings of the $Z$ boson to fermions. Furthermore, we
have $F_S\left(m,m,M\right) = F_{S'}\left(m,M,M\right)$, and the only
contributions in the fourth line in Eq.~\eqref{eq:res:general}
proportional to diagonal $Z$ couplings arise from the finite parts of
the field renormalisation constants.

As before, given the charges of the internal fermions, the charges of
the internal scalars and vectors are determined by the charges of the
external particles, via $Q_{s,v} = 1/3 + Q_f$; so the sums in
Eq.~\eqref{eq:res:general} effectively run only over the particles
with positive charge. Finally, we remind the reader that we suppressed
colour indices throughout; coloured particles can easily be taken into
account with the understanding that the sums also run over all
components of fields in multiplets of $SU(3)$.

\subsubsection{Vector-like quarks}

As an example, we consider vector-like quarks in a representation that
does not generate tree-level FCNC transitions in the down-quark
sector~\cite{delAguila:2000rc}. For simplicity, we treat the case of
an up-type singlet vector-like quark, $U$, with charge $2/3$, isospin
zero, and mass $m_U$. 

The additional quark $U$ mixes with the SM fermions via Yukawa term in
the Lagrangian, given by
\begin{equation}
\mathcal{L}_Y = - \sum_{k} \overline{Q}_{L}^k Y_{k}^U U_{R} \tilde H +
\text{h.c.}\, 
\end{equation}
(note that hard mixing terms $\mathcal{L}_\text{mix} = - \sum_{k=1}^3
m_{uu}^k \, \overline{u}_R^k U_L + \text{h.c.}$ can be eliminated via
a field rotation). Here $k$ is a SM generation index, and the charge
conjugate of the Higgs field $H$ is given by $\tilde H = i \sigma^2
H^* = i \sigma^2 (H^\dagger)^T$.

We obtain the contributions to the $Z$ penguin by inserting the
appropriate couplings into the general result~\eqref{eq:res:general}.
First, note that all RH $Z$ couplings as well as the LH tree-level $Z$
couplings to down-type quarks are not changed from their SM
values. Therefore, we need only the following additional Feynman
rules:
\begin{equation}
\begin{split}
g_{Z \bar u_iu_j}^L = \frac{e}{s_wc_w} \bigg[ \frac{1}{2} (V
  V^\dagger)_{ij} - \frac{2}{3} s_w^2 \delta_{ij} \bigg]\,, \quad
g_{W^+ \bar u_i d_j}^L = \frac{e}{\sqrt{2}s_w} V_{ij}\,, \quad
g_{Z W^+ W^-} = \frac{e c_w}{s_w}
\end{split}
\end{equation}
where $V$ is the generalized, non-unitary CKM matrix (see,
e.g.,~\cite{Branco:1999fs, Ahmady:2001qh}).  (Note that the Higgs
couplings contribute only indirectly to the $Z$ penguin, via the
mixing angles comprised by the matrix $V$.) The resulting expression
for the $Z$ penguin is
\begin{equation}\label{eq:res:vlq}
\begin{split}
C & = V_{td} V_{ts}^* f_1(x_t) + V_{Ud} V_{Us}^* f_1(y_U) + V_{Ud}
V_{ts}^* (V V^\dagger)_{tU} f_4(x_t, y_U) + V_{td} V_{Us}^* (V
V^\dagger)_{Ut} f_4(y_U, x_t) \\ & + V_{td} \big[ V_{cs}^* (V
  V^\dagger)_{ct} + V_{us}^* (V V^\dagger)_{ut} \big] f_2(x_t) +
V_{Ud} \big[ V_{cs}^* (V V^\dagger)_{cU} + V_{us}^* (V V^\dagger)_{uU}
  \big] f_2(y_U) \\ & + V_{ts}^* \big[ V_{cd} (V V^\dagger)_{tc} +
  V_{ud} (V V^\dagger)_{tu} \big] f_3(x_t) + V_{Us}^* \big[ V_{cd} (V
  V^\dagger)_{Uc} + V_{ud} (V V^\dagger)_{Uu} \big] f_3(y_U)
\end{split}
\end{equation}
where $x_t = m_t^2/M_W^2$, $y_U = m_U^2/M_W^2$, and the loop functions
are given by
\begin{equation}
\begin{split}
f_1(x) & = \frac{x(x-6)}{8(x-1)} + \frac{x(2+3x) \log x}{8(x-1)^2} \,,
\quad f_2(x) = \frac{x \log x}{8(1-x)} \,, \quad f_3(x) = - \frac{x}{8}
- f_2(x) \,, \\[2mm]
f_4(x,y) & = \frac{x(x-2y+xy)\log x}{8(x-1)(x-y)} - \frac{(x-1)y^2\log
y}{8(y-1)(x-y)} - \frac{x}{8} \,.
\end{split}
\end{equation}
The box contribution for external neutrinos (needed for the rare $K
\to \pi \nu \bar \nu$ decays) is
\begin{equation}
B = V_{td} V_{ts}^* \, g(x_t) + V_{Ud} V_{Us}^* \, g(y_U)
\end{equation}
where
\begin{equation}
g(x) = \frac{x}{x-1} - \frac{x \log x}{(x-1)^2} \,.
\end{equation}

\subsubsection*{The large-mass limit}

As a simple application we study the limit of a vector-like quark with
a mass much larger than the electroweak symmetry-breaking scale. In
general, we expect the effects of the heavy quark to decouple -- all
contributions to physical observables should be suppressed by a power
of $m_U$. This is not immediately obvious from the loop function; to
see this, it is necessary to expand also the mixing matrix $V$ in
inverse powers of $m_U$. In a basis where the SM up-quark Yukawas are
diagonal, the leading contributions can be written as (see also the
discussion in Ref.~\cite{Branco:1999fs})
\begin{equation}\label{eq:Vexpanded}
V = 
\begin{pmatrix}
V_\text{CKM}\\[2mm]
\tfrac{v}{\sqrt{2} m_U} Y^{U \dagger} V_\text{CKM}
\end{pmatrix}\,,
\end{equation}
where $V_\text{CKM}$ is the SM CKM matrix. Using this explicit form,
it can easily be shown that the non-decoupling terms
in~\eqref{eq:res:vlq} cancel. 

The leading terms in the large-mass expansion are thus of order
$1/m_U^2$, and they can be captured by an effective theory description
(cf. Ref.~\cite{AguilarSaavedra:2008zc}). We will treat here the
special case $Y_1^U = Y_2^U = 0$, so that only the flavor-diagonal
top-quark couplings will be modified, but no FCNC transitions in the
up-quark sector are generated. Working again in a basis where the
up-quark SM Yukawas are diagonal, only the following operators are
generated at order $1/m_U^2$~\cite{delAguila:2000rc}:
\begin{equation}\label{eq:basis_unbroken}
\begin{split}
  Q_{H q,33}^{(3)} & \equiv (H^\dagger i\! \stackrel{\leftrightarrow}{D_\mu^a} H) (\bar Q_{L}^3 \gamma^\mu \sigma^a Q_{L}^3) \,, \\
  Q_{H q,33}^{(1)} & \equiv (H^\dagger i\! \stackrel{\leftrightarrow}{D}_\mu H) (\bar Q_{L}^3 \gamma^\mu Q_{L}^3) \,, \\
  Q_{u H,33} & \equiv (H^\dagger H) (\bar Q_{L}^3 \tilde H t_{R}) \,.
\end{split}
\end{equation}
These operators contain the Higgs doublet $H$ and its charge conjugate
$\tilde H$, the left-handed third-generation quark doublet $Q_{L}^3$,
and the right-handed top quark $t_R$. Moreover, $\sigma^a$ are the
Pauli matrices and $D_\mu$ is the SM gauge-covariant derivative and we
defined
\begin{equation}
\begin{split}
  (H^\dagger i\! \stackrel{\leftrightarrow}{D_\mu} H) & = i
  H^\dagger \big(D_\mu H \big) - i \big(D_\mu H \big)^\dagger 
  H \,, \\
  (H^\dagger i\! \stackrel{\leftrightarrow}{D_\mu^a} H) & = i
  H^\dagger \sigma^a \big(D_\mu H \big) - i \big(D_\mu H
  \big)^\dagger \sigma^a H \,,
\end{split}
\end{equation}
so that the operators $Q_{H q,33}^{(3)}$ and $Q_{H q,33}^{(1)}$ are
manifestly Hermitian. 

Here, we are not interested in Higgs physics observables, so we will
concentrate on the $Z$-penguin operators $Q_{H q,33}^{(3)}$ and $Q_{H
  q,33}^{(1)}$. The implications of constraints from rare meson decays
for anomalous $t \bar t Z$ couplings in the limit of heavy quark
masses have been treated in Ref.~\cite{Brod:2014hsa} in an effective
theory framework. We want to compare these results to those obtained
in our concrete model.

The effective theory approach allows to calculate the
leading-logarithmic contribution to the rare meson decays via operator
mixing (see Ref.~\cite{Brod:2014hsa} for details). From
Ref.~\cite{delAguila:2000rc} we can read off the Wilson coefficients
\begin{equation}
  C_{H q,33}^{(3)} = - C_{H q,33}^{(1)} = \tfrac{1}{4} V_{td} V_{ts}^*
  \big| Y_3^{U}\big|^2 \,,
\end{equation}
while the results in Ref.~\cite{Brod:2014hsa} provide the logarithmic
contribution to the $X$and $Y$ functions
\begin{equation}
\begin{split}
  \delta Y^{\rm NP} = \delta X^{\rm NP} = - \frac{3 + 2 x_t}{2}
  C_{\phi q,33}^{(1)} \frac{v^2}{\Lambda^2} \log \frac{\mu_W}{\Lambda}
  = - \frac{3 + 2 x_t}{8} V_{td} V_{ts}^* \big|Y_3^U \big|^2
  \frac{v^2}{m_U^2} \log \frac{\mu_W}{m_U} \,,
\end{split}
\end{equation}
where $x_t = m_t^2/M_W^2$ and we identified $\Lambda = m_U$. Inserting
the expansion~\eqref{eq:Vexpanded} into the explicit result, we obtain
the logarithmic term
\begin{equation}
\begin{split}
X \big|_{\log} = - \frac{3 + 2 x_t}{8} V_{td} V_{ts}^* \big|
Y_3^{U}\big|^2 \frac{v^2}{m_U^2} \log \frac{M_W}{m_U}
\end{split}
\end{equation}
which reproduces the EFT result. 

In fact, the EFT approach at one-loop allows only to calculate the
leading-logarithmic term in the $1/m_U^2$ expansion. Having the full
theory at hand, we can now check whether this is a reasonable
appoximation. Choosing, as an example, a mass $m_U = 1\,$TeV, and
adding the box contribution to obtain a gauge-invariant result, we
find that the logarithmic term dominates the remaining terms of order
$1/m_U^2$ by a factor of seven for the rare $K \to \pi \nu \bar \nu$
decays, and by a factor of four for $B_s \to \mu^+ \mu^-$. We see that
the EFT result gives, in this instance, a good estimate of the leading
contributions.

Another interesting question is the contribution of dimension-eight
operators (corresponding to terms suppressed by $1/m_U^4$). We find
that the dimension-six terms dominate over the dimension-eight
contributions for masses $m_U \gtrsim 150\,$GeV -- in particular, for
all masses where an expansion in $v/m_U$ is justified.

\subsubsection{Charginos in the MSSM}

As a further check of our formalism, we compare the result for
additional scalars and fermions with the chargino contributions to the
$Z$ penguin in the MSSM.  The particles in the loop are the two
charginos, $(\tilde \chi_1, \tilde \chi_2)$, and the six up-type
squarks, $(\tilde U_j)$, where we follow the notation of
Ref.~\cite{Rosiek:1995kg}.  Using the charge conjugated charginos
$(\tilde \chi_1^c, \tilde \chi_2^c)$ inside the loop, the relevant
coupling constants read
\begin{equation}
  \begin{split}
g_{Z U_i^- U_j^+} &= - \frac{e}{s_{2W}} \left( \sum_{I=1}^3 Z_U^{Ii*} Z_U^{Ij} - \frac{4}{3} s_W^2 \delta_{ij} \right) \,, \\
g^L_{Z \overline{\chi^c_i} \chi^c_j} &= - \frac{e}{s_{2w}} 
\left( Z_-^{1i} Z_-^{1j *} + (c_W^2 - s_W^2) \delta_{ij} \right) \\
g^R_{Z \overline{\chi^c_i} \chi^c_j} &= - \frac{e}{s_{2w}} 
\left( Z_+^{1i} Z_+^{1j *} + (c_W^2 - s_W^2) \delta_{ij} \right) \\
y_{U^-_i\overline{\chi^c_j}d_k}^L &= \sum_{I=1}^3 V_{Ik} \big(- \frac{e}{s_W} Z_U^{Ii*} Z_+^{1j} + Y_{u}^I Z_U^{(I+3)i *} Z_+^{2j} \big) \,,  \\
y_{U^-_i\overline{\chi^c_j}d_k}^R &= - \sum_{I=1}^3 V_{Ik} \big(Y_{d}^k Z_U^{Ii *} Z_-^{2j} \big) \,,  \\
  \end{split}
\end{equation}
Using these coupling constants and the standard model couplings of the
Z-Boson to the down quarks we reproduce the results of the chargino
contributions presented in Ref.~\cite{Buras:2000qz}.

\section{Conclusion}

In this work we presented a manifestly finite result for the
three-point function involving two light SM fermions and the $Z$ boson
(the ``$Z$ penguin''), in generic extensions of the SM that satisfy
the condition of perturbative unitarity. The $Z$ penguin is the main
ingredient for the prediction of decay rates for rare meson decays.

We allow for an arbitrary number of heavy new scalar, fermionic, and
vector particles. The vector particles are assumed to obtain their
mass via the spontaneous breaking of a gauge symmetry. The specific
form of the result is independent of the symmetry-breaking mechanism.

We have derived manifestly finite results for the case of arbitrary
charged internal particles. The results depend on a reduced set of
physical couplings (reflecting the structure of the underlying
symmetry group). This elimination of redundant couplings is important
in particular when one performs a fit to flavor or collider data. The
presentation of explicit results for neutral particles is relegated to
future work. Furthermore, we plan to apply our methods also in the
context of dark matter and collider physics to study simplified models
with a consistent UV behaviour.

\addcontentsline{toc}{section}{Acknowledgements}
\section*{Acknowledgements}
We thank Sandro Casagrande for collaboration at an earlier stage of
this work, and Andrzej Buras for comments on the manuscript. This work
was performed in part at the Aspen Center for Physics, which is
supported by National Science Foundation grant PHY-1066293.  MG is
supported in part by the UK STFC under Consolidated Grant ST/L000431/1
and also acknowledges support from COST Action CA16201 PARTICLEFACE.


\appendix

\section{Result for the box diagrams}
\label{sec:app_box}

The box-diagram contribution to the effective Hamiltonian for $d_j \to
d_i \ell_k \bar \ell_l$ transitions reads
\begin{equation}
\begin{split}
  \label{eq:generic_box_Heff}
  \mathcal H_{\rm eff}^{\Delta F = 1 \, \rm Box} = 
  \frac{4 G_F}{\sqrt{2}} \frac{\alpha}{2 \pi s_w^2}
  \bigg\{ & \sum_{\sigma_1 \sigma_2} \Big(B^{S\sigma_1\sigma_2}_{d_jd_i\ell_l\ell_k}
  (\bar d_i P_{\sigma_1} d_j) (\bar \ell_k P_{\sigma_2} \ell_l)\\
  & \qquad + B^{V\sigma_1\sigma_2}_{d_jd_i\ell_l\ell_k} (\bar d_i \gamma^\mu P_{\sigma_1} d_j)
  (\bar \ell_k \gamma_\mu P_{\sigma_2} \ell_l) \Big)\\
  & + \sum_{\sigma} B^{T\sigma}_{d_jd_i\ell_l\ell_k}
  (\bar d_i \sigma^{\mu\nu} P_{\sigma} d_j)
  (\bar \ell_k \sigma_{\mu\nu} P_{\sigma} \ell_l) \bigg\} \,,
\end{split}
\end{equation}
with $\sigma_{\mu\nu} = i[\gamma^\mu,\gamma^\nu]/2$. The use of chiral
Fierz identities (see \eg \cite{Nishi:2004st}) shows that the
tensor structures with mixed chirality $(\bar d_i \sigma^{\mu\nu}
P_{\sigma} d_j) (\bar \ell_k \sigma_{\mu\nu} P_{\bar\sigma}\, \ell_l)$
vanish identically. Hence, they are omitted from
\eqref{eq:generic_box_Heff}. We decompose the box functions
$B^{M\sigma_1\sigma_2}_{d_jd_i\ell_l\ell_k}$ as
\begin{equation}
\begin{split}
  \label{eq:generic_box_coefficients}
  B^{M\sigma_1\sigma_2}_{d_jd_i\ell_l\ell_k} = \frac{\sqrt{2}}{G_F}
  \frac{s_w^2}{\alpha} \frac{1}{32 \pi} \times \bigg\{
  & \isum{f_1f_2v_1v_2} \Big(
  \tilde c_{f_1f_2v_1v_2}^{M\sigma_1\sigma_2} \,
  \tilde{D}_0\big(m_{f_1},m_{f_2},M_{v_1},M_{v_2}\big)\\
  & \qquad \qquad + c_{f_1f_2v_1v_2}^{M\sigma_1\sigma_2}
  D_0\big(m_{f_1},m_{f_2},M_{v_1},M_{v_2}\big) \Big) \\
  &\quad + \isum{f_1f_2v_1s_1} \Big(
  \tilde c_{f_1f_2v_1s_1}^{M\sigma_1\sigma_2}
  \tilde{D}_0 \big(M_{s_1},m_{f_1},m_{f_2},M_{v_1}\big)\\
  & \qquad \qquad 
  + c_{f_1f_2v_1s_1}^{M\sigma_1\sigma_2}
  D_0\big(M_{s_1},m_{f_1},m_{f_2},M_{v_1}\big) \Big) \\
  &\quad + \isum{f_1f_2s_1s_2} \Big(
  \tilde c_{f_1f_2s_1s_2}^{M\sigma_1\sigma_2}
  \tilde{D}_0\big(M_{s_1},M_{s_2},m_{f_1},m_{f_2}\big)\\
  & \qquad \qquad + 
  c_{f_1f_2s_1s_2}^{M\sigma_1\sigma_2}
  D_0\big(M_{s_1},M_{s_2},m_{f_1},m_{f_2}\big) \Big) \bigg\}
  \,, 
\end{split}
\end{equation}
where the superscript takes the values $M = S,V,T$. The coefficients
for the vector projection are given by
\begin{align} 
  c_{f_1f_2v_1v_2}^{V\sigma_1\sigma_2} &= m_{f_1}^2 m_{f_2}^2
  \bigg(\frac{1}{M_{v_1}^2} + \frac{1}{M_{v_2}^2}\bigg) \,
  g_{\bar{v}_2\bar d_if_1}^{\sigma_1} g_{v_1\bar{f}_1d_j}^{\sigma_1} \bigg(
  g_{\bar{v}_1\bar \ell_kf_2}^{\sigma_2} g_{v_2\bar{f}_2\ell_l}^{\sigma_2}
  - \flipV2 \bigg)
  \,,\notag \displaybreak[0]\\
  \tilde c_{f_1f_2v_1v_2}^{V\sigma_1\sigma_2}
  &= - \frac{1}{4(M_{v_1}^2 + M_{v_2}^2)} c_{f_1f_2v_1v_2}^{V\sigma_1\sigma_2}\\
  & \quad + \bigg( g_{\bar{v}_1\bar \ell_kf_2}^{\sigma_2}
  g_{v_2\bar{f}_2\ell_l}^{\sigma_2} - 4\,\flipV2 \bigg)
  \times \begin{cases}
    - g_{\bar{v}_2\bar d_if_1}^{\sigma_1} g_{v_1\bar{f}_1d_j}^{\sigma_1} ,
    &\mkern-10mu \sigma_1 = \sigma_2 \\
    g_{\bar{v}_1\bar d_if_1}^{\sigma_1} g_{v_2\bar{f}_1d_j}^{\sigma_1} ,
    &\mkern-10mu \sigma_1 \neq \sigma_2 
  \end{cases}
  \,,\notag \displaybreak[0]\\[10pt]
  \label{eq:generic_boxV}
  c_{f_1f_2v_1s_1}^{V\sigma_1\sigma_2} &= m_{f_1} m_{f_2} \Big(
  g_{\bar{v}_1\bar d_if_1}^{\sigma_1} y_{s_1\bar{f}_1d_j}^{\sigma_1}
  + \flipV1 \Big) \Big(
  y_{\bar{s}_1\bar \ell_kf_2}^{\bar \sigma_2} g_{v_1\bar{f}_2\ell_l}^{\sigma_2}
  + \flipV2 \Big)
  \,, \displaybreak[0]\\
  \tilde c_{f_1f_2v_1s_1}^{V\sigma_1\sigma_2} &=
  -\frac{1}{4M_{v_1}^2} c_{f_1f_2v_1s_1}^{V\sigma_1\sigma_2}
   \,,\qquad
  c_{f_1f_2s_1s_2}^{V\sigma_1\sigma_2} = 0
  \,,\notag \displaybreak[0]\\[5pt]
  \tilde c_{f_1f_2s_1s_2}^{V\sigma_1\sigma_2} &= - \frac{1}{4} \,
  y_{\bar{s}_2\bar d_if_1}^{\bar \sigma_1} y_{s_1\bar{f}_1d_j}^{\sigma_1} \bigg(
  y_{\bar{s}_1\bar \ell_kf_2}^{\bar \sigma_2} y_{s_2\bar{f}_2\ell_l}^{\sigma_2}
  - \flipV2 \bigg)
  \,.\notag
\end{align}
Here, $\flipV1$ represents contributions for which we interchange the
coupling constants via $g^\sigma_{\ldots\, \bar d_i f_1} \leftrightarrow
g^\sigma_{\ldots\, \bar f_1 d_j}$ and $y^\sigma_{\ldots\, \bar d_i f_1}
\leftrightarrow y^{\bar \sigma}_{\ldots\, \bar f_1 d_j}$, and
$\flipV2$ acts analogously on $\ell_{l,k}$ and $f_2$. For the scalar
projections we obtain the coefficients
\begin{align} 
  \tilde c_{f_1f_2v_1v_2}^{S\sigma_1\sigma_2}  &= m_{f_1} m_{f_2}
  \bigg(\frac{1}{M_{v_1}^2} + \frac{1}{M_{v_2}^2}\bigg) \,
  g_{\bar{v}_2 \bar d_i f_1}^{\bar \sigma_1} g_{v_1 \bar{f}_1 d_j}^{\sigma_1}
  \Big\{ g_{\bar{v}_1 \bar \ell_k f_2}^{\bar \sigma_2}
  g_{v_2 \bar{f}_2 \ell_l}^{\sigma_2} + \flipS2 \Big\}
  \,,\notag \displaybreak[0]\\
  c_{f_1f_2v_1v_2}^{S\sigma_1\sigma_2} &= - \frac{4 M_{v_1}^2
    M_{v_2}^2 + m_{f_1}^2 m_{f_2}^2}{M_{v_1}^2 + M_{v_2}^2} \tilde
  c_{f_1f_2v_1v_2}^{S\sigma_1\sigma_2}
  \,,\notag \displaybreak[0]\\
  \tilde c_{f_1f_2v_1s_1}^{S\sigma_1\sigma_2} &= \Big(
  y_{\bar{s}_1 \bar d_i f_1}^{\sigma_1} g_{v_1 \bar{f}_1 d_j}^{\sigma_1}
  - \flipS1 \Big) \Big(
  g_{\bar{v}_1 \bar \ell_k' f_2}^{\bar \sigma_2} y_{s_1 \bar{f}_2 l}^{\sigma_2}
  - \flipS2 \Big)
  \,,\label{eq:generic_boxS} \displaybreak[0]\\
  c_{f_1f_2v_1s_1}^{S\sigma_1\sigma_2} &= 
  - \frac{m_{f_1}^2 m_{f_2}^2}{M_{v_1}^2}
  \tilde c_{f_1f_2v_1s_1}^{S\sigma_1\sigma_2} \,,\qquad
  \tilde c_{f_1f_2s_1s_2}^{S\sigma_1\sigma_2} = 0
  \,,\notag \displaybreak[0]\\
  c_{f_1f_2s_1s_2}^{S\sigma_1\sigma_2} &= - m_{f_1} m_{f_2} \,
  y_{\bar s_2 \bar d_i f_1}^{\sigma_1} y_{s_1 \bar f_1 d_j}^{\sigma_1} \Big(
  y_{\bar s_1 \bar \ell_k f_2}^{\sigma_2} y_{s_2 \bar f_2 \ell_l}^{\sigma_2}
  + \flipS2 \Big) \,,\notag
\end{align}
and finally the tensor coefficients read
\begin{align}
  c_{f_1f_2v_1v_2}^{T\sigma} &= - m_{f_1} m_{f_2}
  g_{\bar v_2 \bar d_i f_1}^{\bar \sigma} g_{v_1 \bar f_1 d_j}^{\sigma}
  \Big( g_{\bar v_1 \bar \ell_k f_2}^{\bar \sigma}
  g_{v_2 \bar f_2 \ell_l}^{\sigma} - \flipS2 \Big)
  \,,\notag \displaybreak[0]\\
  \tilde c_{f_1f_2v_1v_2}^{T\sigma} &= - \frac{1}{4}
  \bigg(\frac{1}{M_{v_1}^2} + \frac{1}{M_{v_2}^2}\bigg)
  c_{f_1f_2v_1v_2}^{T\sigma}
  \,,\label{eq:generic_boxT} \displaybreak[0]\\
  \tilde c_{f_1f_2v_1s_1}^{T\sigma} &= \frac{1}{4} \Big(
  y_{\bar s_1 \bar d_i f_1}^{\sigma} g_{v_1 \bar f_1 d_j}^{\sigma}
  + \flipS1 \Big) \Big(
  g_{\bar v_1 \bar \ell_k f_2}^{\bar \sigma} y_{s_1 \bar f_2 \ell_l}^{\sigma}
  + \flipS2 \Big)
  \,,\notag \displaybreak[0]\\
  c_{f_1f_2v_1s_1}^{T\sigma} &= c_{f_1f_2s_1s_2}^{T\sigma}
  = \tilde c_{f_1f_2s_1s_2}^{T\sigma} = 0
  \,.\notag
\end{align}
Similar to above $\flipS1$ stands for additional
contributions, here with the coupling constants
$y^\sigma_{\; \cdot \; \bar d_i f_1} \leftrightarrow
y^{ \sigma}_{\; \cdot \; \bar f_1 d_j}$ and
$g^\sigma_{\; \cdot \; \bar d_i f_1} \leftrightarrow
g^{\bar\sigma}_{\; \cdot \; \bar f_1 d_j}$ interchanged, and $\flipS2$ acts
analogously on $\ell_{l,k}$ and $f_2$.

It is possible to extract the generic effective Hamiltonian for the
$|\Delta F| = 2$ boxes from the above results, if one keeps in mind
that the external particle-antiparticle pairs allow for additional
Wick contractions of the transition amplitude and that the
conventional normalization of the effective Hamiltonian is different.
We use Fierz identities to express the result in terms of the operator
basis of~\cite{Gorbahn:2009pp}, where
\begin{align}
  & \mathcal H_\textrm{eff}^{\Delta F=2 \; \textrm{Box}} = 
  \frac{G_F^2}{4 \pi^2} M_W^2 \bigg( \sum_{\sigma}
  S^{V\sigma\sigma}_{1,d_jd_i} \, Q_{1,d_id_j}^{V\sigma\sigma} + \sum_{n=1,2}
  \Big( S^{LR}_{n,d_jd_i} \, Q_{n,d_id_j}^{LR} + \sum_{\sigma}
  S^{S\sigma\sigma}_{n,d_jd_i} \, Q_{n,d_id_j}^{S\sigma\sigma} \Big) \bigg)
  \,,\notag\\[5pt]
  &\mkern+80mu \begin{aligned}
    Q_1^{V\sigma\sigma} &= (\bar d_j \gamma_{\mu} P_\sigma d_i)
        (\bar d_j \gamma^{\mu} P_\sigma d_i) \,,\\
    Q_1^{LR} &= (\bar d_j \gamma_{\mu} P_L d_i)
        (\bar d_j \gamma^{\mu} P_R d_i) \,,\;
    & Q_2^{LR} &= (\bar d_j P_L d_i) (\bar d_j P_R d_i) \,,\\
    Q_1^{S\sigma\sigma} &= (\bar d_j P_\sigma d_i)
        (\bar d_j P_\sigma d_i) \,,\;
    &Q_2^{S\sigma\sigma} &= (\bar d_j^\alpha P_\sigma d_i^\beta)
        (\bar d_j^\beta P_\sigma d_i^\alpha) \,.
  \end{aligned}
  \label{eq:generic_Heff_DeltaF2_Ham}
\end{align}
A summation over colour indices $\alpha,\beta$ is understood. In order
to obtain the coefficients $S$ from \eqref{eq:generic_boxV} --
\eqref{eq:generic_boxT}, the following prescription holds:
\begin{equation}
  \label{eq:generic_DeltaS2_coefficient_rules}
  \begin{aligned}
  S^{V\sigma\sigma}_{1,d_jd_i} &= 4\,
  B^{V\sigma\sigma}_{d_jd_i\ell_l\ell_k}|_{\ell_{k,l} \,\to\, d_{i,j}} \,,\\
  S^{LR}_{1,d_jd_i} &= 8\, B^{VLR}_{d_jd_i\ell_l\ell_k}
  |_{\ell_{k,l} \,\to\, d_{i,j}} \,,\;
  & S^{LR}_{2,d_jd_i} &= 8\, B^{SLR}_{d_jd_i\ell_l\ell_k}
  |_{\ell_{k,l} \,\to\, d_{i,j}} \,,\\
  S^{S\sigma\sigma}_{1,d_jd_i} &= \big( 4\,
  B^{S\sigma\sigma}_{d_jd_i\ell_l\ell_k} - 16\,
  B^{T\sigma\sigma}_{d_jd_i\ell_l\ell_k}\big)
  |_{\ell_{k,l} \,\to\, d_{i,j}} \,,\;
  & S^{S\sigma\sigma}_{2,d_jd_i} &= -32\,
  B^{T\sigma\sigma}_{d_jd_i\ell_l\ell_k}|_{\ell_{k,l} \,\to\, d_{i,j}} \,.
  \end{aligned}
\end{equation}

\section{Loop functions}\label{sec:loopfunctions}

In the UV-divergent loop functions we set 
$\epsilon = (4-D)/2$. The loop functions are defined
as (cf.~\cite{Gorbahn:2009pp})
\begin{equation}
\begin{split}
  \frac{i}{(4 \pi)^2} B_0 \left( m_1, m_2 \right)
  \left( \frac{4 \pi}{\mu^2} e^{-\gamma_E} \right)^\epsilon 
  &=
  \int \frac{d^D q}{(2 \pi)^D}
  \frac{1}{q^2-m_1^2} \frac{1}{q^2-m_2^2} \\
  \frac{i}{(4 \pi)^2} C_0 \left( m_1, m_2, m_3 \right)
  \left( \frac{4 \pi}{\mu^2} e^{-\gamma_E} \right)^\epsilon 
  &=
  \int \frac{d^D q}{(2 \pi)^D}
  \frac{1}{q^2-m_1^2} \frac{1}{q^2-m_2^2} \frac{1}{q^2-m_3^2} \\
  \frac{i}{(4 \pi)^2} D_0 \left( m_1, m_2, m_3, m_4 \right)
  \left( \frac{4 \pi}{\mu^2} e^{-\gamma_E} \right)^\epsilon 
  &=
  \int \frac{d^D q}{(2 \pi)^D}
  \frac{1}{q^2-m_1^2} \frac{1}{q^2-m_2^2} \frac{1}{q^2-m_3^2} 
  \frac{1}{q^2-m_4^2} 
\end{split}
\label{eq:13}
\end{equation}
These functions read: 
\begin{equation}
\label{eq:loop-expl}
\begin{split}
  B_0 \left( m_1, m_2 \right) &=
  \frac{1}{\epsilon} + 1 + 
  \frac{m_1^2 \log \left(\frac{\mu^2}{m_1^2}\right)
    - m_2^2 \log \left(\frac{\mu^2}{m_2^2}\right)}{
    m_1^2-m_2^2} \\
  C_0 \left( m_1, m_2, m_3 \right) &= \frac{
    m_1^2 \, m_2^2 \log \left(\frac{m_2^2}{m_1^2}\right)+
    m_3^2 \, m_2^2 \log \left(\frac{m_3^2}{m_2^2}\right)+
    m_1^2 \, m_3^2 \log \left(\frac{m_1^2}{m_3^2}\right)}{
    \left(m_1^2-m_2^2\right)
    \left(m_1^2-m_3^2\right)
    \left(m_2^2-m_3^2\right)} \\
  D_0 \left( m_1, m_2, m_3, m_4 \right) &= \sum_\textrm{cyclic permutations}
\frac{m_1^2 \log m_1^2}{\big(m_1^2-m_2^2\big) \big(m_1^2-m_3^2\big)
  \big(m_4^2-m_1^2\big)}
\end{split}
\end{equation}
\begin{equation}
\begin{split}\label{eq:relation}
\tilde C_0 (m_1,m_2,m_3) =&
B_0 (m_2,m_3) + m_1^2 C_0(m_1,m_2,m_3) \\
\tilde D_0 (m_1,m_2,m_3,m_4) =&
C_0 (m_2,m_3,m_4) + 
m_1^2 D_0(m_1,m_2,m_3,m_4) 
\end{split}
\end{equation}

\section{Complete list of STIs for Feynman Rules}
\label{sec:other_STIs}

Here we list the full set of identities for the four-point couplings
provided by the tree-level analysis of STIs described in
section \ref{sec:generic_lagrangian}. Their derivation involves the
following additional three- and four-point coupling
\begin{equation} \label{eq:add_lagrangian}
\begin{split}
    \mathcal{L}_{\rm 3 \& 4} &\supset 
    - {\textstyle\sum\limits_{v_i,v_j}} \Big\{
    \tfrac{i e}{2} \, \omega_{A, v_i v_j} F^{\mu\nu} V_{v_i,\mu}^a V_{v_j,\nu}^a
    + \tfrac{i g_s}{2} \, \omega_{G, v_i v_j}^{abc} G^{a, \mu\nu}
    V_{v_i,\mu}^b V_{v_j,\nu}^c \Big\} \\
    &+
    \tfrac{1}{6} \isum{s_1 s_2 s_3 s_4} \mk50 
    g_{s_1s_2s_3}^{abc} \Big( h_{s_1}^a h_{s_2}^b h_{s_3}^c \Big) +
    \tfrac{1}{24} \isum{s_1 s_2 s_3 s_4} \mk50 
    g_{s_1s_2s_3s_4}^{abcd} \Big( h_{s_1}^a h_{s_2}^b h_{s_3}^c h_{s_4}^d \Big) \\
    &+
    \tfrac{1}{8} \isum{v_1 v_2 v_3 v_4} \mk50 
    g_{v_1v_2v_3v_4}^{abcd} \Big( V_{v_1,\mu}^a V_{v_2}^{b,\mu} V_{v_3,\nu}^c V_{v_4}^{d,\nu} \Big) +
    \tfrac{1}{4} \isum{v_1 v_2 s_1 s_2} \mk50 
    g_{v_1v_2s_1s_2}^{abcd} \Big( V_{v_1,\mu}^a V_{v_2}^{b,\mu}  h_{s_1}^c h_{s_2}^d \Big) \,. 
\end{split}
\end{equation}
The kinetic term of the vector fields $V_{v_i}$ and the couplings to
the field strength tensors $\omega$ contribute to triple gauge boson
vertices with one photon or gluon. $U(1)_\text{em} \times
SU(3)_\text{colour}$ gauge invariance restricts these couplings to be
of the form
\begin{equation}
  \omega_{A, ij} = \delta_{\mk15 \ol{i} j} Q_{V_j} \,,\qquad
  \omega_{G, ij}^{abc} = \delta_{\mk15 \ol{i} j} T^a_{V_j, bc} \,.
\end{equation}

The quartic interactions involving unphysical Goldstone couplings can
be expressed in terms of physical couplings as follows:
\begin{equation}
  \label{eq:STI_bosons_Goldstone}
  \begin{aligned}
    & g_{\p_1\p_2\p_3\p_4} = \tfrac{- \sigma_{v_1} \sigma_{v_2}
      \sigma_{v_3} \sigma_{v_4}}{4 m_{v_1} m_{v_2} m_{v_3} m_{v_4}} \,
    \isum{s_5} \, m_{s_5}^2 \Big( g_{v_1v_4s_5} g_{v_2v_3 \bar s_5 } +
    \text{symm}(v_2,v_3,v_4) \Big) \,,\\[5pt]
    & g_{v_1v_2\p_3\p_4} = \tfrac{\sigma_{v_3} \sigma_{v_4}}{2 m_{v_3}
      m_{v_4}} \bigg\{ \isum{s_5} \, g_{v_1v_2s_5} g_{v_3v_4 \bar s_5 } \\
    &\mkern+155mu + \isum{v_5} \Big( \big(m_{v_1}^2 + m_{v_2}^2 - 2
    m_{v_5}^2\big) \big(g_{v_1v_4v_5} g_{v_2v_3 \bar v_5 } +
    g_{v_1v_3v_5} g_{v_2v_4 \bar v_5 } \big) \\
    &\mkern+205mu - \tfrac{(m_{v_1}^2 - m_{v_2}^2)(m_{v_3}^2 -
      m_{v_4}^2)}{m_{v_5}^2} \, g_{v_1v_2v_5} g_{v_3v_4 \bar v_5 } \Big)
    \bigg\} \,,\\[5pt]
    & g_{\p_1\p_2\p_3s_4} = \tfrac{i\, \sigma_{v_1}
      \sigma_{v_2} \sigma_{v_3}}{6 m_{v_1} m_{v_2} m_{v_3}}
    \bigg\{ \isum{s_5} \, ( 2 m_{s_4}^2 - 3 m_{s_5}^2 )
    g_{v_1s_4s_5} g_{v_2v_3 \bar s_5 } \\
    &\mkern+180mu + m_{s_4}^2 \isum{v_5} \, \tfrac{m_{v_2}^2 -
      m_{v_3}^2}{m_{v_5}^2} \, g_{v_1v_5s_4} g_{v_2v_3 \bar v_5 } +
    \text{symm}(v_1,v_2,v_3) \bigg\} \,,\\[5pt]
    & g_{v_1v_2\p_3s_4} = \tfrac{i\, \sigma_{v_3}}{m_{v_3}} \bigg\{
    \isum{v_5} \Big( - g_{v_1v_5s_4} g_{v_2v_3 \bar v_5 } - g_{v_2v_5s_4}
    g_{v_1v_3 \bar v_5 } \\
    & \mkern+158mu + g_{v_3v_5s_4} g_{v_1v_2 \bar v_5 }
    \tfrac{m_{v_1}^2 - m_{v_2}^2}{2 m_{v_5}^2} \Big)
    + \isum{s_5} \, g_{v_3s_4s_5} g_{v_1v_2 \bar s_5 } \bigg\} \,,\\[5pt]
    & g_{\p_1\p_2s_3s_4} = \tfrac{\sigma_{v_1} \sigma_{v_2}}{2 m_{v_1}
      m_{v_2}} \bigg\{ \isum{s_5} \, \Big( g_{s_3s_4s_5} g_{v_1v_2 \bar s_5 }
    \\
    &\mkern+185mu + \Big[ g_{v_1s_3s_5} g_{v_2s_4 \bar s_5 } \, (2
    m_{s_5}^2 - m_{s_3}^2 - m_{s_4}^2) + \text{symm}(s_3,s_4) \Big] \Big)
    \\
    &\mkern+120mu + \isum{v_5} \,\Big(g_{v_5s_3s_4} g_{v_1v_2 \bar v_5 }
    \, \big( m_{s_3}^2 - m_{s_4}^2 \big) \, \tfrac{m_{v_1}^2 -
      m_{v_2}^2}{m_{v_5}^2} \\
    &\mkern+170mu + \Big[ g_{v_1v_5s_4} g_{v_2 \bar v_5 s_3} \, \big(
    m_{s_3}^2 + m_{s_4}^2 \big) \, \tfrac{1}{4 m_{v_5}^2} +
    \text{symm}(v_1,v_2) \Big]\Big) \bigg\} \,,\\[5pt]
    & g_{\p_1s_2s_3s_4} = \tfrac{i\, \sigma_{v_1}}{m_{v_1}} \bigg\{
    \isum{s_5} \, g_{s_2s_3s_5} g_{v_1s_4 \bar s_5 } - \isum{v_5} \,
    g_{v_5s_2s_3} g_{v_1 \bar v_5 s_4} \tfrac{m_{s_2}^2 - m_{s_3}^2}{2
      m_{v_5}^2} + \text{symm}(s_2,s_3,s_4) \bigg\} \,.
  \end{aligned} \raisetag{23pt}
\end{equation}
Furthermore, all quartic interactions with vectors are fixed by
the three-point couplings
\begin{equation}
  \label{eq:STI_bosons_quartic}
  \begin{aligned}
    & g_{v_1v_2v_3v_4} = \isum{v_5} \, \Big( g_{v_1v_4v_5}
    g_{v_2v_3 \bar v_5 } + g_{v_1v_3v_5} g_{v_2v_4 \bar v_5 } \Big) \,, \\
    & g_{v_1v_2s_3s_4} = \bigg\{ \isum{v_5} \, g_{v_1v_5s_4}
    g_{v_2 \bar v_5 s_3} \tfrac{1}{4m_{v_5}^2} - \isum{s_5} \,
    g_{v_1s_3s_5} g_{v_2s_4 \bar s_5 } + \text{symm}(v_1,v_2) \bigg\} \,.
  \end{aligned} \raisetag{23pt}
\end{equation}
The remaining, purely bosonic sum-rules are given by
\begin{equation}
  \label{eq:STI_bosons_sumrules}
  \begin{aligned}
    & \isum{s_5}\, \Big( g_{v_1v_2s_5} g_{v_3v_4 \bar s_5 } -
    g_{v_1v_4s_5} g_{v_2v_3 \bar s_5 } \Big) \\
    &\mkern+100mu = \isum{v_5} \bigg( g_{v_1v_3v_5} g_{v_2v_4 \bar v_5 } \,
    \big( 2 m_{v_5}^2 - m_{v_1}^2 - m_{v_2}^2 - m_{v_3}^2 - m_{v_4}^2
    \big) \\
    &\mkern+160mu + g_{v_1v_2v_5} g_{v_3v_4 \bar v_5 } \Big( m_{v_5}^2
    + \tfrac{(m_{v_1}^2 - m_{v_2}^2) (m_{v_3}^2 - m_{v_4}^2)}{m_{v_5}^2}
    \Big) \\
    &\mkern+160mu + g_{v_1v_4v_5} g_{v_2v_3 \bar v_5 } \Big( m_{v_5}^2
    - \tfrac{(m_{v_1}^2 - m_{v_4}^2) (m_{v_2}^2 - m_{v_3}^2)}{m_{v_5}^2}
    \Big) \bigg) \,, \\[5pt]
    & \isum{s_5}\, \Big( g_{v_1v_2s_5} g_{v_3s_4 \bar s_5 } -
    g_{v_2v_3s_5} g_{v_1s_4 \bar s_5 } \Big) \\
    &\mkern+100mu = \isum{v_5} \bigg( g_{v_2v_5s_4} g_{v_1v_3 \bar v_5 } +
    g_{v_3v_5s_4} g_{v_1v_2 \bar v_5 } \,\tfrac12 \Big( 1 -
    \tfrac{m_{v_1}^2 - m_{v_2}^2}{m_{v_5}^2} \Big) \\
    &\mkern+250mu + g_{v_1v_5s_4} g_{v_2v_3 \bar v_5 } \,\tfrac12 \Big( 1
    + \tfrac{m_{v_2}^2 - m_{v_3}^2}{m_{v_5}^2} \Big) \bigg) \,, \\[5pt]
    & \isum{v_5} \, g_{v_1v_2 \bar v_5 } g_{v_5s_3s_4} = \isum{v_5} \,
    \tfrac{1}{4 m_{v_5}^2} \Big( g_{v_1v_5s_4} g_{v_2 \bar v_5 s_3} -
    g_{v_1v_5s_3} g_{v_2 \bar v_5 s_4} \Big) \\
    &\mkern+135mu + \isum{s_5}\, \Big( g_{v_1s_3s_5} g_{v_2s_4 \bar s_5 }
    - g_{v_2s_3s_5} g_{v_1s_4 \bar s_5 } \Big) \,.
  \end{aligned}
\end{equation}
All rules given here and in section \ref{sec:generic_lagrangian} hold
in the limiting case of any of the particles being massless. The
couplings of photons or gluons of course simplify considerably and can
be expressed by using the $U(1)$ and $SU(3)$ charges defined in
\eqref{eq:generic_covariant_derivative}. Moreover they can be
separated into two classes: Couplings involving $V$-$A$ or $V$-$G$
transitions (the generalization of $Z$-$A$ transitions known from the
SM) and those without. Couplings of the latter class are either
directly included in the covariant kinetic Lagrangian or zero. This
includes\footnote{To derive the first line, one has to note that $g_{A
v_1 s_2}$ vanishes. This coupling has to come from the kinetic term of
a multiplet $\Phi$ of the full gauge group, and is thus proportional
to $e Q_\Phi A_\mu V_a^\mu (\Phi^\dagger T_{V_a} \Phi)$. However the
Goldstone directions are precisely given by $T_{V_a} \langle \Phi
\rangle$.}
\begin{equation}
  \label{eq:qed_couplings_no_mixing}
  \begin{gathered}
    g_{A \p_1 \p_2} = g_{A A \p_1 \p_2} = g_{A \p_1 s_1}
    = g_{A A \p_1 s_1} = 0 \,, \\
    g_{A A \p_1 \p_2} = -2 i ( g_{A v_1  \bar v_1 } )^2
    \delta_{v_1  \bar v_2 } \,.
  \end{gathered}
\end{equation}
The equations also hold with $G$ instead of $A$. The class of
couplings with $V$-$A$ or $V$-$G$ transitions has to be derived from
the STIs. They are given by
\begin{equation}
  \begin{split}
    & g_{A v_1 \p_2} = i e \, m_{v_1} \delta_{v_1  \bar v_2 } \,, \\
    & g_{A v_1 \p_2 \p_3} = e \, \sigma_{v_2} \sigma_{v_3} \big(
    Q_{v_2} - Q_{v_3} \big) \, \tfrac{m_{v_2}^2 + m_{v_3}^2 -
      m_{v_1}^2}{2 m_{v_2} m_{v_3}} \, g_{v_1 v_2 v_3} \,, \\
    & g_{A v_1 \p_2 s_3} = -i e \, \sigma_{v_2} \big( Q_{v_2} -
    Q_{s_3} \big) \, \tfrac{1}{2 m_{v_2}} \, g_{v_1 v_2 s_3} \,, \\
    & g_{A v_1 v_2 v_3} = - e \, ( Q_{v_2} - Q_{v_3} ) \, g_{v_1 v_2
      v_3} \,, \\
    & g_{A v_1 s_2 s_3} = e \, ( Q_{s_2} - Q_{s_3} ) \, g_{v_1 s_2
      s_3} \,. \\
  \end{split}
\end{equation}
In the quartic coupling we used charge conservation to simplify the
right hand side. The corresponding equations with a gluon are
\begin{equation}
  \begin{split}
    & g_{G v_1 \p_2}^{c_3 c_1 c_2} = i \sigma_{v_2} \big(
    T_{v_2}^{c_3} \big)_{c_1 c_2} \, m_{v_1} \delta_{v_1  \bar v_2 } \,, \\
    & g_{G v_1 \p_2 \p_3}^{c_4 c_1 c_2 c_3} = \sigma_{v_2}
    \sigma_{v_3} \tfrac{m_{v_2}^2 + m_{v_3}^2 - m_{v_1}^2}{2 m_{v_2}
      m_{v_3}} \, \isum{c_5} \, \Big( \big( T_{v_2}^{c_4} \big)_{c_2
      c_5} \, g_{v_1 v_2 v_3}^{c_1 c_5 c_3} - \big( T_{V_{a_3}}^{c_4}
    \big)_{c_3 c_5} \, g_{v_1 v_2 v_3}^{c_1 c_2 c_5} \Big) \,, \\
    & g_{G v_1 \p_2 s_3}^{c_4 c_1 c_2 c_3} = -i \, \sigma_{v_2} \,
    \tfrac{1}{2 m_{v_2}} \, \isum{c_5} \, \Big( \big( T_{v_2}^{c_4}
    \big)_{c_2 c_5} \, g_{v_1 v_2 s_3}^{c_1 c_5 c_3} - \big( T_{s_3}^{c_4}
    \big)_{c_3 c_5} \, g_{v_1 v_2 s_3}^{c_1 c_2 c_5} \Big) \,, \\
    & g_{G v_1 v_2 v_3}^{c_4 c_1 c_2 c_3} = - \isum{c_5} \, \Big( \big(
    T_{v_2}^{c_4} \big)_{c_2 c_5} \, g_{v_1 v_2 v_3}^{c_1 c_5 c_3} - \big(
    T_{v_3}^{c_4} \big)_{c_3 c_5} \, g_{v_1 v_2 v_3}^{c_1 c_2 c_5} \Big)
    \,, \\
    & g_{G v_1 s_1 s_2}^{c_4 c_1 c_2 c_3} = \isum{c_5} \, \Big( \big(
    T_{s_1}^{c_4} \big)_{c_2 c_5} \, g_{v_1 s_1 s_2}^{c_1 c_5 c_3} - \big(
    T_{s_2}^{c_4} \big)_{c_3 c_5} \, g_{v_1 s_1 s_2}^{c_1 c_2 c_5 }
    \Big)\,. \\
  \end{split}
\end{equation}
We have checked explicitly that the STIs for five-point vertex
functions do not imply additional sum rules.

\addcontentsline{toc}{section}{References}
\bibliography{literature}
\bibliographystyle{./bibstyle_generic}

\end{document}